\documentstyle[aaspp4,11pt]{article}

\received{2002 March 31}
\accepted{2002 May 22}

\slugcomment{accepted for publication in ApJ}

\begin{document}

\title
{The Effect of the Hall Term on the Nonlinear Evolution of the
Magnetorotational Instability: II. Saturation Level and Critical
Magnetic Reynolds Number}

\author
{Takayoshi Sano and James M. Stone}
\affil
{Department of Astronomy, University of Maryland, College Park, MD
20742-2421; sano@astro.umd.edu}

\begin{abstract}
The nonlinear evolution of the magnetorotational instability (MRI) in
weakly ionized accretion disks, including the effect of the Hall term
and ohmic dissipation, is investigated using local three-dimensional 
MHD simulations
and various initial magnetic field geometries.  When the magnetic
Reynolds number, $Re_{M} \equiv v_{\rm A}^2 / \eta \Omega$ (where
$v_{\rm A}$ is the Alfv{\'e}n speed, $\eta$ the magnetic diffusivity,
and $\Omega$ the angular frequency), is initially larger than a
critical value $Re_{M,{\rm crit}}$, the MRI evolves into MHD turbulence
in which angular momentum is transported efficiently by the Maxwell
stress.  If $Re_{M} < Re_{M,{\rm crit}}$, however, ohmic dissipation
suppresses the MRI, and the stress is reduced by several orders of
magnitude.  The critical value is in the range of 1 -- 30 depending on
the initial field configuration.  The Hall effect does not modify the
critical magnetic Reynolds number by much, but enhances the saturation
level of the Maxwell stress by a factor of a few.  We show that the
saturation level of the MRI is characterized by $v_{{\rm A}z}^2 / \eta
\Omega$, where $v_{{\rm A}z}$ is the Alfv{\'e}n speed in the nonlinear
regime along the vertical component of the field.  The condition for
turbulence and significant transport is given by $v_{{\rm A}z}^2 / \eta 
\Omega \gtrsim 1$, and this critical value is independent of the
strength and geometry of the magnetic field or the size of the Hall
term.  If the magnetic field strength in an accretion disk can be
estimated observationally, and the magnetic Reynolds number 
$v_{\rm A}^2 / \eta \Omega$ is larger than about 30, this would imply
the MRI is operating in the disk.
\end{abstract}


\section{INTRODUCTION}

The nonlinear regime of the magnetorotational instability (MRI) (Balbus
\& Hawley 1991) can strongly affect the structure and evolution of
accretion disks.  In ideal MHD, the MRI initiates and sustains MHD
turbulence in which angular momentum is transported outward by Maxwell
(magnetic) stress.  Thus, the MRI is thought to be the most promising
source of anomalous viscosity in disks.  In weakly ionized disks,
however, the coupling between the gas and magnetic field may be so poor
that nonideal MHD effects must be considered.

When nonthermal processes (such as irradiation by cosmic rays or high
energy photons) dominate the ionization rate in the disk, the abundance
of charged particles decreases as the number density of the neutral gas
$n_n$ increases.  At
high densities ($n_n \gtrsim 10^{18}$ cm$^{-3}$), ohmic dissipation
dominates the
evolution of the MRI (Jin 1996; Fleming, Stone, \& Hawley 2000; Sano \&
Inutsuka 2001).  At low densities ($n_n \lesssim 10^{13}$ cm$^{-3}$), 
ambipolar diffusion dominates (Blaes \& Balbus 1994;
Hawley \& Stone 1998).  However, at intermediate densities, the ions
are decoupled from the magnetic field and can drift relative to the
electrons (which remain frozen-in to the field).  Thus, in this regime
Hall currents can significantly alter the MHD of the plasma, and the
Hall term dominates the other nonideal MHD effects.  Detailed
calculations reveal that the Hall term could be important in dwarf nova
disks in quiescence, and in protoplanetary disks around young stellar
objects (Sano \& Stone 2002, hereafter Paper I).

The properties of the MRI are strongly affected by the Hall term
(Wardle 1999; Balbus \& Terquem 2001).  In Hall MHD the critical
wavenumber and maximum growth rate of the MRI both depend on the
direction of the magnetic field with respect to the angular frequency
vector {\boldmath $\Omega$}.  The Hall term can increase the maximum
growth rate when the disk is threaded by a uniform vertical field in
the same direction as {\boldmath $\Omega$}, whereas the MRI can be
completely suppressed if the field is oppositely directed to {\boldmath
$\Omega$}.

In Paper I, the effect of the Hall term on the nonlinear evolution of
the MRI was investigated using axisymmetric numerical MHD simulations.
These calculations included ohmic dissipation as well as the Hall
effect, because at some densities both processes may be important.  In
two dimensions (2D), depending on the relative amplitude of the Hall and
ohmic dissipation terms in the induction equation, the MRI evolves into
either a two-channel flow without saturation, or MHD turbulence that
eventually dies away.

In this paper, we continue our study of the Hall effect on the
the MRI using fully three-dimensional (3D) numerical MHD simulations.
Previous studies
have shown that only in 3D is sustained MHD turbulence generated by the
MRI (e.g., Hawley, Gammie, \& Balbus 1995).  Moreover, the effect of
nonaxisymmetric modes on the saturation amplitude and resulting stress
can only be explored in 3D.  Previous 3D simulations including only
ohmic dissipation have shown that there exists a critical value for the
magnetic Reynolds number $Re_{M,{\rm crit}}$ for significant turbulence
and stress to be generated in the saturated state (Fleming
et al. 2000; Sano \& Inutsuka 2001).  Moreover, this critical value
depends on the field geometry in the disk (Fleming et al. 2000).  The
value of $Re_{M,{\rm crit}}$ has important implications for the
structure and evolution of accretion disks (Gammie 1996; Glassgold,
Najita, \& Igea 1997; Sano et al. 2000).  For example, the possibility
of layered accretion and the size of the putative dead zone in
protoplanetary disks depends on the critical value $Re_{M,{\rm crit}}$
(Fleming \& Stone 2002).  In addition to studying the properties of the
nonlinear regime of the MRI in 3D Hall MHD, an important goal of this
paper is to investigate whether the inclusion of the Hall term
significantly modifies the value of $Re_{M,{\rm crit}}$.

In a weakly ionized plasma, the induction equation should include terms
which represent three non-ideal MHD effects: the Hall effect, ohmic
dissipation, and ambipolar diffusion.  The importance of these terms
relative to the inductive term is determined mainly by the magnitude of
the neutral density, the ionization fraction, and the field strength.  In
paper I, we calculated the ratios of these terms to the inductive term by
solving the ionization equilibrium equations in both dwarf nova and
protoplanetary disks.  At the typical density of a dwarf nova disk ($n_n
\sim 10^{18}$ cm$^{-3}$), the ambipolar diffusion term is much smaller
than the inductive term, while the Hall and ohmic dissipation terms are
equally important and of order the inductive term.  These two terms are
also important in the inner, dense regions of protoplanetary disks. Thus,
to study the MRI in these systems, we solve the induction equation
including the Hall and ohmic dissipation terms, but neglecting the
ambipolar diffusion term.

Several definitions of the magnetic Reynolds number are possible; in
this paper we define $Re_{M} \equiv v_{\rm A}^2 / \eta \Omega$, where
$v_{\rm A}$ is the Alfv{\'e}n speed, $\eta$ is the magnetic
diffusivity, and $\Omega$ is the angular frequency.  This definition
uses the wavelength of the fastest growing mode of the MRI as the
typical length scale, i.e. $L = v_{\rm A} / \Omega$, and therefore
directly captures the effect of resistivity on the linear dispersion
relation (Jin 1996; Sano \& Miyama 1999), as well as the local
properties of the saturated state (Paper I).  Note, however, that with
this definition $Re_{M}$ depends on the magnetic field strength.
Moreover, it is different from the magnetic Reynolds number $Re_{M}' =
c_s^2 / \eta \Omega$ used in Fleming et al. (2000) by a factor $v_{\rm
A}^2 / c_s^2$, where $c_s$ is the sound speed.

The plan of this paper is as follows.  In \S 2, the numerical method
and initial conditions used in the calculations are described.  The
results from simulations using an initially uniform vertical field are
discussed in \S 3, from simulations using an initially vertical field
with zero-net flux in \S 4, and from simulations using an initially
uniform toroidal field in \S 5.  The criteria for significant angular
momentum transport, and the application of the results to protoplanetary
disks, are discussed in \S 6, and our results are summarized in \S 7.

\section{NUMERICAL METHOD}

We adopt the local shearing box approximation (Hawley et al. 1995) for our
calculations.  The MHD equations are written in a local Cartesian frame
of reference ($x$, $y$, $z$) corotating with the disk at angular
frequency $\Omega$, where $x$ is oriented in the radial direction, $y$
is in the azimuthal direction, and $z$ is in the vertical direction.
Vertical gravity is ignored in this analysis, since our computational
domain represents a region much smaller than the thickness of the
disk.  The gas is assumed to be partially ionized, and composed of
ions, electrons, and neutrals.  Charge neutrality is assumed, so that
$n_i = n_e$, where $n_i$ and $n_e$ are the number density of ions and
electrons, respectively.  The equations we solve are then
\begin{equation}
\frac{\partial \rho}{\partial t} + 
\mbox{\boldmath $v$} \cdot \nabla \rho = 
- \rho \nabla \cdot \mbox{\boldmath $v$} ~,
\end{equation}
\begin{equation}
\frac{\partial \mbox{\boldmath $v$}}{\partial t} + 
\mbox{\boldmath $v$} \cdot \nabla \mbox{\boldmath $v$} = 
- \frac{\nabla P}{\rho} 
+ \frac{\mbox{\boldmath $J$} \times \mbox{\boldmath $B$}}{c \rho} 
- 2 \mbox{\boldmath $\Omega$} \times \mbox{\boldmath $v$} 
+ 2 q \Omega^2 x \hat{\mbox{\boldmath $x$}} ~,
\label{eqn:eom}
\end{equation}
\begin{equation}
\frac{\partial \epsilon}{\partial t} + 
\mbox{\boldmath $v$} \cdot \nabla \epsilon = 
- \frac{P \nabla \cdot \mbox{\boldmath $v$}}{\rho} 
+ \frac{4 \pi \eta \mbox{\boldmath $J$}^2}{c^2 \rho} ~,
\end{equation}
\begin{equation}
\frac{\partial \mbox{\boldmath $B$}}{\partial t} = 
\nabla \times 
\left( \mbox{\boldmath $v$} \times \mbox{\boldmath $B$} 
- \frac{4 \pi \eta \mbox{\boldmath $J$}}{c} 
- \frac{\mbox{\boldmath $J$} \times \mbox{\boldmath $B$}}{e n_e} \right) 
~,
\label{eqn:ineq}
\end{equation}
where {\boldmath $v$} is the neutral velocity, $\epsilon$ is the
specific internal energy,
\begin{equation}
\mbox{\boldmath $J$} = 
\frac{c}{4 \pi} \left( \nabla \times \mbox{\boldmath $B$} \right)
\end{equation}
is the current density, and $c$ is the speed of light.
The equation of motion (eq. [\ref{eqn:eom}]) includes the Coriolis force,
$-2 \mbox{\boldmath $\Omega$} \times \mbox{\boldmath $v$}$, and the
tidal expansion of the effective potential, $2 q \Omega^2 x$, with
$q = 3 / 2$ for a Keplerian disk. 
The pressure is given by $P = ( \gamma - 1 ) \rho \epsilon$ with $\gamma
= 5 / 3$.
The last two terms in the induction equation (\ref{eqn:ineq}) represent
ohmic dissipation and the Hall effect, respectively, where $\eta$ is the
magnetic diffusivity and $e$ the elementary electric charge.

These equations are solved using a finite-difference code (Sano,
Inutsuka, \& Miyama 1999).  The hydrodynamics module of our scheme is
based on the second order Godunov scheme (van Leer 1979) using a
nonlinear Riemann solver modified to account for the effect of
tangential magnetic fields.  The evolution of magnetic fields is
calculated with the constrained transport (CT) method (Evans \& Hawley
1988), which guarantees the divergence free condition, $\nabla \cdot
\mbox{\boldmath $B$} = 0$, is satisfied throughout the calculation. 
Each term of the electromotive force 
in the induction equation (\ref{eqn:ineq}) is
solved by an operator split
solution procedure (see Paper I).

The initial physical quantities are assumed to be spatially uniform
($\rho = \rho_0$ and $P = P_0$) except for the Keplerian shear flow
$v_{y0} = - q \Omega x$.  The initial magnetic field is very weak for
all models, so that radial force balance in the initial state is
between Coriolis and tidal forces.  Three initial magnetic
field configurations are considered in this paper; a uniform vertical
field $B_z = B_0$, a zero-net flux vertical field $B_z(x) = B_0 \sin (
2 \pi x / L_x)$, where $L_x$ is the size of the computational domain in
the radial direction, and a uniform toroidal field $B_y = B_0$.

The calculations are performed in a local volume bounded by $x = \pm
H/2$, $y = \pm 2H$, and $z = \pm H/2$, where $H \equiv ( 2 / \gamma
)^{1/2} c_{s0} / \Omega$ is the scale height of the disk and $c_{s0}$ is 
the initial sound speed.  Most of
the runs use a standard grid resolution of $32 \times 128 \times 32$
uniform zones.  In the azimuthal and vertical directions, periodic
boundary conditions are used.  For the radial boundary, a sheared
periodic boundary condition (Hawley \& Balbus 1992) is adopted.  The
magnetic flux within the shearing box must be conserved unless a net
flux of radial field exists (Hawley et al. 1995).  However, we have
found that numerical errors in the net flux of the vertical and
azimuthal fields are non-negligible (more than 10 \% in 10 orbits) when
the EMF in the ghost zones at the radial boundary are constructed from
applying the sheared periodic boundary condition to the magnetic field
and velocity.  Instead, if we apply the the sheared periodic boundary
condition to the azimuthal component of the EMF, the error in the
vertical flux is reduced to $\sim  0.1$ \%.

We choose the normalizations $\rho_0 = 1$, $H = 1$, and $\Omega =
10^{-3}$.
Then the sound speed and gas pressure are initially $2 c_{s0}^2 /
\gamma = 10^{-6}$ and $P_0 = 5 \times 10^{-7}$, respectively.
Initial perturbations are introduced as spatially uncorrelated pressure
and velocity fluctuations.  
These fluctuations have a zero mean value with a maximum amplitude of
$| \delta P | / P_0 = 10^{-2}$ and $| \delta \mbox{\boldmath $v$} | /
c_{s0} = 10^{-2}$.

Our calculations are characterized by three dimensionless parameters.
The first is the ratio of the gas to initial magnetic pressure,
\begin{equation}
\beta_0 = \frac{8 \pi P_0}{B^2_0} = \frac{2 c_{s0}^2}{\gamma v_{A0}^2} ~,
\end{equation}
where $v_{A0} = B_0/(4 \pi \rho_0)^{1/2}$ is the initial Alfv{\'e}n speed.
This parameter measures the initial field strength.
The second is the magnetic Reynolds number, which is defined as
\begin{equation}
 Re_{M0} = \frac{v_{A0}^2}{\eta\Omega} ~.
\end{equation}
We assume the magnetic diffusivity $\eta$ is constant in our
calculations.  This parameter measures the importance of ohmic dissipation.
Finally, the third
parameter is 
\begin{equation}
X_0 = \frac{cB_0\Omega}{2 \pi e n_{e0} v_{A0}^2} ~,
\label{eqn:x}
\end{equation}
where $n_{e0}$ is the initial number density of electrons.
We assume the electron abundance is constant throughout our calculations,
thus $n_e = n_{e0} \rho / \rho_0$.
The value of $X_0$ measures the importance of the Hall term; note that
the sign of $X_0$ depends on the direction of the magnetic field.

\section{SIMULATIONS WITH A UNIFORM VERTICAL FIELD}

Simulations that begin with a uniform vertical field are listed in
Tables \ref{tbl:z} -- \ref{tbl:zl}.  The critical wavelength
$\lambda_{\rm crit}$ and the maximum growth rate $\sigma_{\max}$
are obtained from the dispersion relation derived by Balbus \&
Terquem (2001) for axisymmetric perturbations.  Disturbances with a
wavelength longer than $\lambda_{\rm crit}$ are unstable to the MRI.
The critical wavelength is of the order of $2 \pi v_{\rm A} / \Omega$
in most cases when ohmic dissipation is inefficient ($Re_{M0} >
1$).  However, when the Hall parameter is just below zero ($-4 < X_0 <
0$), $\lambda_{\rm crit} \rightarrow 0$ and small-scale perturbations
become unstable to the MRI.  Ohmic dissipation, on the other hand, can
stabilize small-scale perturbations, and thus make $\lambda_{\rm crit}$
longer and $\sigma_{\max}$ smaller when $Re_{M0} < 1$.  In 2D, when $X_0
< 0$ or $Re_{M0} < 1$ saturation of the instability occurs, whereas
exponential growth of a two-channel flow without saturation persists
for other models (Paper I).  In 3D without the Hall term, the nonlinear
evolution of the MRI is characterized by recurrent appearance of
two-channel flow (Fleming et al. 2000; Sano \& Inutsuka 2001).  We now
examine how this evolution is changed by the inclusion of the Hall
effect.

In addition to model parameters, Tables \ref{tbl:z} -- \ref{tbl:zl}
contain several fundamental quantities that are referred to in the
following discussion.
Hereafter, the single brackets $\langle f \rangle$ imply a volume average 
of quantity $f$, whereas double brackets $\langle \negthinspace \langle
f \rangle \negthinspace \rangle$ denotes a time and volume average. 
If not otherwise stated, the time average is taken over the last 10
orbits of the calculation for the models in this section.

\subsection{The Fiducial Models}

We describe three models in
detail: models Z2 ($X_0 = 2$), Z3 ($X_0 = 0$), and Z4 ($X_0 = -2$).
These fiducial models have the same initial field strength and
magnetic Reynolds number, $\beta_0 = 3200$ and $Re_{M0} = 100$.
The only difference between the non-zero $X_0$ models (Z2 and Z4), is
the direction of the vertical field.
Ohmic dissipation is so weak in these models that the
evolution is dominated by the Hall effect.

Figure \ref{fig:em-t-z} shows the time evolution of the volume-averaged
magnetic energy $\langle B^2 / 8 \pi \rangle / P_0$ for these models,
where time is measured in orbits $t_{\rm rot} = 2 \pi / \Omega$.
During the linear phase of the MRI, magnetic energy is amplified
exponentially.  As expected from the linear analysis, the $X_0 = 2$
model has a larger growth rate than the $X_0 = - 2$ run initially.  In
the nonlinear regime, MHD turbulence is sustained in all three models.
The saturated magnetic energy shows large fluctuations whose amplitude
depends on the size of the Hall parameter:  as $X_0$ decreases, the
amplitude of the variability becomes smaller.  The saturation level of
the magnetic energy is higher
in the positive $X_0$ model than in the negative
$X_0$ model.

The efficiency of angular momentum transport is given by the
turbulent stress,
\begin{equation}
w_{xy} = - \frac{B_x B_y}{4 \pi} + \rho v_x \delta v_y ~,
\end{equation}
where the first and second terms are the Maxwell stress ($w_M \equiv -
B_x B_y / 4 \pi$) and Reynolds stress ($w_R \equiv \rho v_x \delta
v_y$), respectively.
The total stress $w_{xy}$ is related to the $\alpha$ parameter of
Shakura \& Sunyaev (1973) by $\alpha = w_{xy}/P_0$.
Figure \ref{fig:wm-t-z} shows the time evolution of the Maxwell stress
normalized by the initial pressure $P_0$ for the fiducial models.
The time-averaged Maxwell and Reynolds stresses are listed in Table
\ref{tbl:z}.
In MHD turbulence generated by the MRI, the Maxwell stress is a few
times larger than the Reynolds stress.  Large time variability can be
seen in the evolution of the stress, with the amplitude of fluctuations
in the $X_0 = 2$ run about $\Delta \langle w_M \rangle / P_0 \sim 0.4$,
which is much larger than the time-averaged value of the stress
$\langle \negthinspace \langle w_M \rangle \negthinspace \rangle / P_0
= 0.16$.  In the model with negative $X_0$, on the other hand, the
variation is much smaller throughout the evolution.

The spike-shaped variations in the magnetic energy and stress correspond
to the recurrent appearance and breakup of the two-channel flow.  These
same variations could be seen 
in the 3D simulations of the MRI without the Hall effect studied by Sano \&
Inutsuka (2001).
Figure \ref{fig:z2} shows images of the magnetic energy distribution 
in the $X_0 = 2$ run (Z2) at 18, 21, and 25 orbits.
The velocity field is also shown by arrows.
The top and bottom panel are chosen at times near the peak of
spikes in the magnetic energy, whereas the middle panel is near a minimum.
The large-scale structure of the flow in the top and bottom panels clearly
shows the axisymmetric two-channel flow.
The spatial dispersion in the magnetic pressure is very high during these
phases; at 18
orbits $\langle \delta P_{\rm mag}^2 \rangle^{1/2} / \langle P_{\rm
mag} \rangle = 1.5$.
The magnetic pressure is comparable to the gas pressure
($\langle P_{\rm mag} \rangle / \langle P \rangle = 0.65$), and the gas
pressure and density have large spatial dispersions
($\langle \delta P^2 \rangle^{1/2} / \langle P \rangle = 0.52$ and
$\langle \delta \rho^2 \rangle^{1/2} / \langle \rho \rangle = 0.38$).

The axisymmetric channel flow is unstable to nonaxisymmetric modes of
the parasitic instability (Goodman \& Xu 1994).
After the breakup of the channel flow, organized large-scale structure
disappears and the disk is occupied by disorganized MHD turbulence
(Fig. \ref{fig:z2}, {\it middle panel}).
The magnetic energy at this phase is an order of magnitude smaller than
at the peak of the spikes, so that $\langle P\rangle / \langle P_{\rm mag}  
\rangle = 15$ at 21 orbits.
The magnetic pressure is still spatially highly fluctuating, $\langle
\delta P_{\rm mag}^2 \rangle^{1/2} / \langle P_{\rm mag} \rangle = 1.1$, 
but this fluctuation is small relative to the gas pressure,
$\langle \delta P_{\rm mag}^2 \rangle^{1/2} / \langle P \rangle =
0.075$. 
Since the density fluctuation decreases as the fluctuation in the
magnetic pressure relative to the gas pressure $\langle \delta P_{\rm
mag}^2 \rangle^{1/2} / \langle P \rangle$ decreases (Turner, Stone, \&
Sano 2002), the spatial dispersion in the gas pressure and density is
small at this phase; $\langle \delta P^2 \rangle^{1/2} / \langle P
\rangle = 0.14$ and $\langle \delta\rho^2 \rangle^{1/2} / \langle \rho
\rangle = 0.085$.

The snapshots of the magnetic energy in the negative $X_0$ model
(Z4) is shown in Figure \ref{fig:z4}.  In contrast to the $X_0 = 2$ run (Z2)
shown in Figure \ref{fig:z2}, this model has little time variability.
To allow direct comparison, the contour levels  are the same in both figures.
The magnetic energy is typically comparable to or smaller than that during
low energy phase of the $X_0 = 2$ run.
The emergence of the channel flow cannot be seen, and
disorganized MHD turbulence is sustained throughout the evolution.
Because the critical wavenumber $k_{\rm crit} \rightarrow \infty$ in
this case, small-scale disturbances are noticeable in all directions.
The gas pressure is always much larger than the magnetic pressure and
the ratio of their time average is $\langle \negthinspace \langle P
\rangle \negthinspace \rangle / \langle \negthinspace \langle P_{\rm
mag} \rangle \negthinspace \rangle = 32$.
The spatial dispersion in the magnetic energy is very large, $\langle
\delta P_{\rm mag}^2 \rangle^{1/2} / \langle P_{\rm mag} \rangle = 1.5$,
but the gas pressure and density are almost uniformly distributed,
$\langle \delta P^2 \rangle^{1/2} / \langle P \rangle = 0.093$ and
$\langle \delta \rho^2 \rangle^{1/2} / \langle \rho \rangle = 0.054$.
These quantities are quite similar to those during the low magnetic energy
phases of the $X_0 = 2$ run. 

To compare the characteristics of MHD turbulence in the fiducial
models, Table
\ref{tbl:z2} lists a number of
time- and volume-averaged quantities.
Although the magnetic and kinetic energy in the positive $X_0$ run are
about 4 times larger than those in the negative $X_0$ run, interestingly
the turbulence of three fiducial runs has many similarities.
For example, the magnetic and perturbed kinetic energy are almost equal
for all the models, with the ratio $\langle \negthinspace \langle 
\rho \delta v^2 / 2 \rangle \negthinspace \rangle / \langle
\negthinspace \langle B^2 / 8 \pi \rangle \negthinspace \rangle \approx
0.4$.
Because of the shear motion, the azimuthal component of the magnetic
field is amplified most efficiently.
Each component of the magnetic field energy has a similar ratio for all
cases, $\langle \negthinspace \langle B_x^2 \rangle \negthinspace
\rangle : \langle \negthinspace \langle B_y^2 \rangle \negthinspace
\rangle : \langle \negthinspace \langle B_z^2 \rangle \negthinspace
\rangle \approx 3 : 20 : 1$. 
The perturbed kinetic energy is slightly anisotropic, $\langle
\negthinspace \langle \rho v_x^2 \rangle \negthinspace \rangle : \langle
\negthinspace \langle \rho \delta v_y^2 \rangle \negthinspace \rangle :
\langle \negthinspace \langle \rho v_z^2 \rangle \negthinspace \rangle
\approx 3 : 2 : 1$.

The model parameters, $\beta_0$, $Re_{M0}$, and $X_0$, are defined by
the initial state, but they are variable in time.
The time-averaged values of these parameters in the fiducial models are
listed in Table \ref{tbl:z2}.
The magnetic pressure is smaller than the gas pressure even in the
nonlinear stage, so that the time- and volume-averaged plasma beta
$\langle \negthinspace \langle \beta \rangle \negthinspace \rangle
\equiv \langle \negthinspace \langle P / (B^2/ 8 \pi) \rangle
\negthinspace \rangle$ is of the order of 100.
However, the gas pressure is increasing linearly throughout the
evolution due to the dissipation of magnetic field.
Then, the plasma beta $\langle \negthinspace \langle \beta \rangle
\negthinspace \rangle$ must be increasing in time unless some cooling
processes are included.
When the magnetic field is amplified, the magnetic Reynolds number
becomes larger than the initial value.
The time-averaged value is $\langle \negthinspace \langle Re_{M} \rangle
\negthinspace \rangle \equiv \langle \negthinspace \langle v_{\rm A}^2 /
\eta \Omega \rangle \negthinspace \rangle > 10^4$ for the fiducial
models, so that the ohmic dissipation at the nonlinear stage is less
efficient.
We define the effective Hall parameter in the nonlinear regime as 
\begin{equation}
X_{\rm eff} \equiv \frac{c B_z \Omega}{2 \pi e n_e v_{\rm A}^2} 
= \frac{2 c \rho B_z \Omega}{e n_e B^2} ~.
\label{eqn:xeff}
\end{equation}
Because the Hall parameter is inversely proportional to the field
strength, the volume-averaged value $| \langle X_{\rm eff} \rangle |$
decreases as the field is amplified.  
The time-averaged value is $\langle \negthinspace \langle X_{\rm eff}
\rangle \negthinspace \rangle = 0.003$ and $-0.03$ for the $X_0 = 2$ and
$-2$ run, respectively.
Therefore, the effect of the Hall term may be reduced in the
nonlinear regime for these cases.

\subsection{Saturation Level}

In this subsection, we explore in more detail the influence of the model
parameters on
the time- and volume-averaged
stress in the saturated state.
Time-averaged Maxwell and Reynolds stress are listed in Table
\ref{tbl:z} -- \ref{tbl:zl} for all models.
As shown by the fiducial models, the stress exhibits large time
variability in the turbulent state. 
For all the models, the time averaging is taken over the last 10 orbits
of the calculation, which is longer than the typical timescale of
variations in the stress.
The time-averaged quantities taken over the last 40 orbits of the
fiducial models (Z2 and Z4) are similar to the 10 orbit averages, and
the difference in the Maxwell and Reynolds stress is less than 10 \%.
The time-averaged magnetic Reynolds number defined by the vertical
magnetic field $\langle \negthinspace \langle v_{{\rm A}z}^2 / \eta
\Omega \rangle \negthinspace \rangle$ is also listed in the tables.
The meaning of $\langle \negthinspace \langle v_{{\rm A}z}^2 / \eta
\Omega \rangle \negthinspace \rangle$ is discussed later in \S 6.1.

\subsubsection{Effect of Initial Field Strength}

First we study the effect of the initial field strength.
Figure \ref{fig:wm-x-rm1-z} depicts the saturated Maxwell stress as a
function of the Hall parameter $X_0$ for various models with different
$\beta_0$.
All the models shown in this figure has the same magnetic Reynolds
number $Re_{M0} = 1$.
For any $\beta_0$, the saturation levels in the positive $X_0$ runs are
higher than those in the negative $X_0$ runs.
The ratio of the stress between the $X_0 = \pm 2$ runs is 28, 4.1, and
2.6 for $\beta_0 = 800$, 3200, and 12800, respectively.
All models with $X_0 \ge 0$ show large time variability due to the
nonlinear growth of the channel flow, while no growth of the two-channel
flow can be seen in all models with $X_0 < 0$.
If we compare models with the same $X_0$, larger $\beta_0$ models
have a lower saturation level, which means that the magnetic energy and
stress increase as the initial field strength increases, as in the
ideal MHD cases (Hawley et al. 1995). 

Since the linear growth rate of the MRI is higher for $X_0 > 0$ (Balbus
\& Terquem 2000; Paper I), this may account for the higher saturation
level in this case.  In addition, the evolution of the MRI shows the
recurrent growth of the channel flow when the Hall parameter is $X_0
\ge 1$.  Since the channel flow can amplify the magnetic field more
efficiently than disorganized MHD turbulence, this could also be a
reason for the larger saturation level in the positive $X_0$ runs.
This result may also be understood in terms of the linear properties of
the MRI:  If $X_0 \ge 0$, the critical wavelength is proportional to
the field strength $\lambda_{\rm crit} \sim v_{\rm A} / \Omega$, so
that $\lambda_{\rm crit}$ increases as the magnetic energy is
amplified, leading to the emergence of large-scale channel flows.  When
the Hall parameter is negative, on the other hand, the critical
wavenumber for the MRI is infinity, so that small-scale perturbations
are unstable.  The MRI continuously excites small-scale disturbances in
this case, and these small fluctuations impede the nonlinear growth of
the two-channel flow.

\subsubsection{Effect of Magnetic Reynolds Number}

When the magnetic Reynolds number is very small, ohmic dissipation
can dramatically reduce the linear growth rate (Jin 1996) and the
nonlinear saturation level of the MRI (Sano \& Inutsuka 2001).
The dependence of the saturated stress on the magnetic Reynolds number
is illustrated in Figure \ref{fig:wm-x-b32-z}, which shows the
time-averaged stress for the models with $Re_{M0} = 100$, 1, and 0.1.
The same field strength is used for all the models in this figure
($\beta_0 = 3200$).
We find that the positive $X_0$ runs always have a larger stress than
the negative $X_0$ runs.
The differences are by a factor of 4 -- 6 in the $Re_{M0} = 100$ and 1
runs.
For very resistive models with $Re_{M0} = 0.1$, the ratio of the stress
between the $X_0 = 4$ and $-2$ runs is about 100.
However the stress is of the order of $10^{-5}$ -- $10^{-4}$, and this is
more than 2 orders of magnitude smaller than the less resistive cases
($Re_{M0} \ge 1$).
This suggests that strong dissipation can weaken the turbulence
even when the Hall term is included.

\subsubsection{Effect of Resolution and Box Size}

Table \ref{tbl:zh} lists the model parameters and saturation levels for
high resolution runs designed to study the effect of numerical resolution.
Except for the grid resolution, the model parameters of Z2H and Z4H are
the same as those of the non-zero $X_0$ fiducial models Z2 and Z4.
The high resolution models are followed only to 10 orbits, so that
the time average is taken over the last 5 orbits.
In the standard resolution models (Z2 and Z4), the time-averaged stress
over every 5 orbits after 10 orbits has up to 70 \% difference compared
to the average through the last 40 orbits.
It may be that the time-averaged stress in the high resolution models has
uncertainty of the similar size.
We find that most of the nonlinear features shown by the fiducial models
are independent of the grid resolution.
The magnetic energy and stress in the $X_0 = 2$ run (Z2H) is larger
than those in the $X_0 = -2$ run (Z4H) by a factor of about 4.
The saturation levels in models Z2H and Z4H are close to those in the
standard resolution cases.
The positive $X_0$ run shows the nonlinear growth of the channel
flow, and the negative $X_0$ run evolves into disorganized MHD turbulence 
with many small-scale fluctuations.

Table \ref{tbl:zl} lists the models computed with a larger computational box
than the fiducial models, namely $2H \times 8H
\times 2H$.
The  grid resolution is $64 \times 256 \times 64$, so that the grid spacing
is the same as the standard models listed in Table
\ref{tbl:z}.
The other parameters of models Z2L and Z4L are the same as the fiducial
models Z2 and Z4.
As in the standard box size cases, the magnetic energy and stress in the
positive $X_0$ run (Z2L) is larger than those in the negative $X_0$ run
(Z4L) throughout the evolution.
The frequency of spike-shaped variations in the stress is reduced
with a larger box size, and thus disorganized MHD turbulence lasts
most of the time.
The stress in the large box model with $X_0 = 2$ (Z2L) is comparable to
that in the standard model Z2, but the $X_0 = -2$ model (Z4L) is twice
as large as the standard model Z4.
Although the saturation level for models with a uniform vertical
field may be proportional to the box size for ideal MHD simulations
(Hawley et al. 1995), our sample of simulations is too small to
confirm this dependence in this study.

\subsection{Characteristics of Saturated MHD Turbulence}

It is of interest to study the properties of the MHD turbulence driven
by the MRI in Hall MHD.  Figure \ref{fig:spd}a and \ref{fig:spd}b show
the Fourier power spectra of the magnetic energy along the $k_x$,
$k_y$, and $k_z$ axes at 25 orbits for the fiducial models Z2 ($X_0 =
2$) and Z4 ($X_0 = -2$), respectively.  The spectra are averaged over
10 snapshots within 0.1 orbits.  In the $X_0 = 2$ run, the large-scale
channel flow is growing at that time.  Thus, smaller wavenumbers have
larger power, especially in $k_z$, and the power at the inertial range
declines as $k^{-4}$, similar to the ideal MHD case (Hawley et al.
1995).  
For the $X_0 = -2$ run (Fig. \ref{fig:spd}b), the
amplitude of the power is smaller than the $X_0 = 2$ run, although the
shape of the spectra is quite similar to the positive $X_0$ run.  The
slope of the power is slightly gentler in the $X_0 = -2$ run, probably
because modes with larger $k$ are unstable to the MRI in this case.
The power spectra at low magnetic energy phase of the $X = 2$ run (which
is dominated by disorganized MHD turbulence) are similar to
those of the negative $X_0$ model.

When ohmic dissipation is inefficient $Re_{M0} > 1$, the MHD turbulence
generated by the MRI appears to have the characteristic properties.
For example, the Maxwell stress is always proportional to the magnetic
pressure,
\begin{equation}
\langle \negthinspace \langle w_M \rangle \negthinspace \rangle
= (0.455 \pm 0.023) 
\langle \negthinspace \langle P_{\rm mag} \rangle \negthinspace \rangle ~,
\label{eqn:wmpm}
\end{equation}
where the average and the standard deviation are taken from the 17 models 
with $Re_{M0} \ge 1$ listed in Table 1.
Note that this average includes models both with and without the Hall
effect. 
The deviation of this ratio is extremely small, and this is independent
on $X_0$. 
The relation given by equation (\ref{eqn:wmpm}) is the same as in the
ideal MHD cases (Hawley et al. 1995). 
For the models with $Re_{M0} \ge 1$, the ratios between each component
of the magnetic and perturbed kinetic energy ($\langle \negthinspace
\langle B_x^2
\rangle \negthinspace \rangle : \langle \negthinspace \langle B_y^2
\rangle \negthinspace \rangle : \langle \negthinspace \langle B_z^2
\rangle \negthinspace \rangle$ and $\langle \negthinspace \langle \rho
v_x^2 \rangle \negthinspace \rangle : \langle \negthinspace \langle \rho
\delta v_y^2 \rangle \negthinspace \rangle : \langle \negthinspace
\langle \rho v_z^2 \rangle \negthinspace \rangle$) take quite similar
values to the fiducial models.
These ratios have also no correlation with $X_0$.
The average ratio of the Maxwell and Reynolds stress is 
\begin{equation}
\langle \negthinspace \langle w_M \rangle \negthinspace \rangle
= (4.49 \pm 1.66)
\langle \negthinspace \langle w_R \rangle \negthinspace \rangle ~.
\end{equation}
The ratio $\langle \negthinspace \langle w_M \rangle \negthinspace
\rangle / \langle \negthinspace \langle w_R \rangle \negthinspace
\rangle$ is slightly dependent of $X_0$, and the average is $5.77$ and
$3.03$ for the $X_0 = 2$ and $-2$ runs, respectively.

When $Re_{M0} < 1$, the turbulence in the nonlinear regime is weakened
by ohmic dissipation.  In this case, the power spectra of the
turbulence is quite different compared to the $Re_{M0} \ge 1$ runs.
Figure \ref{fig:spd2} shows the power spectrum for model Z13 with
$\beta_0 = 3200$, $Re_{M0} = 0.1$, and $X_0 = 0$.  The saturation level
of the Maxwell stress is very low for this model ($\langle
\negthinspace \langle w_M \rangle \negthinspace \rangle / P_0 = 3.1
\times 10^{-4}$).  Other models with $Re_{M0} = 0.1$ including the Hall
term (models Z11, Z12, and Z14) show little difference in the energy
spectra compared to the $X_0 = 0$ run.  The typical diffusion scale
defined as $k_{\rm diff} \equiv \eta / v_{\rm A0}$, as well as the
critical wavenumber for the MRI, $k_{\rm crit}$, is shown on the plot
of the power spectra of the magnetic energy (Fig. \ref{fig:spd2}, {\it
left panel}).  The diffusion length is close to the critical
wavelength in this case.  For scales smaller than the diffusion length,
fluctuations in the magnetic field dissipate faster than the Alfv{\'e}n
timescale.  Therefore, the spectrum shows a very steep decline and the
slope is proportional to $k^{-8}$ in the dissipation regime $k \gtrsim
k_{\rm diff}$.  The right panel of Figure \ref{fig:spd2} shows the
power spectra of the perturbed kinetic energy, which also is a steeply
decreasing function of $k$.

We find that the energy distribution of the turbulence in the $Re_{M0}
= 0.1$ runs is also quite different from that in the less resistive
models.  In Table \ref{tbl:z2}, time- and volume-averaged quantities
for the $Re_{M0} = 0.1$ runs (Z12, Z13, and Z14) are listed.
The saturated level of the magnetic energy is much lower than the
$Re_{M0} = 100$ runs.  The magnetic field is amplified by the MRI
during the linear phase, but dies away due to the ohmic dissipation, so
that in the nonlinear regime the magnetic energy returns to its initial
value.  Because the net flux through the computational volume is
conserved, the magnetic energy can never completely die away.  The
energy in the perturbed velocity is larger than in the magnetic field,
$\langle \negthinspace \langle \rho \delta v^2 / 2 \rangle
\negthinspace \rangle \gg \langle \negthinspace \langle B^2 / 8 \pi
\rangle \negthinspace \rangle$, and the Reynolds stress is a few times
larger than the Maxwell stress for the $Re_{M0} = 0.1$ models.  The
large kinetic energy is a remnant of the linear growth of the MRI.
Although the perturbed magnetic field
generated by the growth of the MRI is efficiently dissipated in these
models by the large resistivity, the perturbed kinetic energy remains
large due to the small viscosity.
Therefore, if the disk is linearly unstable and the magnetic Reynolds
number is small ($Re_{M0} < 1$), the Reynolds stress could dominate the
Maxwell stress but the efficiency of angular momentum transport
is very small ($\alpha \sim 10^{-5}$ -- $10^{-3}$).

\subsection{Critical Magnetic Reynolds Number}

In this subsection, we consider the critical value of the magnetic
Reynolds number for significant angular momentum transport in accretion
disks with uniform vertical fields (as shown by previous studies, this
critical value depends on the initial field geometry, Fleming et al.
2000).  The dependence of the saturation level
on the magnetic Reynolds number $Re_{M0}$ is shown by Figure
\ref{fig:wm-rm-z}.  Open circles and triangles denotes the results
without the Hall term ($X_0 = 0$) for the model with $\beta_0 = 3200$
and 12800, respectively.  When $Re_{M0} \ge 1$, the saturation level is
of the order of 0.01 and almost independent of the magnetic Reynolds
number.  As shown by Figure \ref{fig:wm-x-rm1-z}, the stress is larger
when the initial field is stronger.  If the magnetic Reynolds number is
less than unity, on the other hand, the saturation level is reduced by
a large factor.  For the $Re_{M0} = 0.1$ runs, the stress is about
$10^{-4}$ for both cases.  Thus, the criterion for significant
turbulence is $Re_{M0} \gtrsim 1$ independent of the field strength.
Note that the critical value of the magnetic diffusivity $\eta$ depends
on the field strength, because the magnetic Reynolds number is a
function of the Alfv{\'e}n speed ($\eta_{\rm crit} = v_{A0}^2 / \Omega
Re_{M0,{\rm crit}} \sim v_{A0}^2 / \Omega$).  

Next the effect of the Hall term on the critical magnetic Reynolds
number is considered.
Filled circles and triangles show the saturated stress in the models
with non-zero Hall parameter $X_0 = 4$ and $-2$, respectively.
Although the stress in the $X_0 = 4$ runs is always several times larger
than that in the $X_0 = - 2$ runs, the dependence on the magnetic
Reynolds number is the same as in the models without the Hall term.
That is, the saturation level is almost independent of $Re_{M0}$ when
$Re_{M0} \ge 1$, but drops significantly as the magnetic Reynolds number
decreases if $Re_{M0} < 1$.

In 2D simulations, all the models with $Re_{M0} < 1$ undergo rapid
decay of the magnetic energy both with and without the Hall term (Paper
I).
Thus, the dependence of the saturation level on the ohmic dissipation
and the Hall effect is consistent with the 2D results.
When $Re_{M0} \ge 1$, the magnetic Reynolds number $\langle
\negthinspace \langle v_{\rm A}^2 / \eta \Omega \rangle \negthinspace
\rangle$ at the nonlinear stage is larger than the initial value, because the
magnetic field is amplified by the MRI.
If $Re_{M0} < 1$, the magnetic energy is unchanged by the instability,
and thus $\langle
\negthinspace \langle v_{\rm A}^2 / \eta \Omega \rangle \negthinspace
\rangle$ remains less than unity.
Therefore, the critical value $\langle
\negthinspace \langle v_{\rm A}^2 / \eta \Omega \rangle \negthinspace
\rangle \sim 1$ is valid
even in the nonlinear regime for uniform $B_z$ models. 

\section{SIMULATIONS WITH A ZERO-NET FLUX VERTICAL FIELD}

We next consider simulations that begin with a zero-net flux vertical
field, that is $B_z(x) = B_0 \sin (2 \pi x / L_x)$ where $B_0$ is a
positive constant.  For this case, the Hall parameter is given by $X(x)
= X_0/\sin (2 \pi x / L_x)$, where $X_0 = 2 c \rho_0 \Omega / e n_{e0}
B_0$.  Thus, the region $x < 0$ has positive $X$ while the region $x >
0$ has negative $X$, and the minimum of the absolute value $|X(x)|$ is
$X_0$.  Table \ref{tbl:s} lists the models computed with a zero-net
flux $B_z$.  The initial plasma beta $\beta_0$, the critical wavelength
$\lambda_{\rm crit}$, and the maximum growth rate $\sigma_{\max}$ in
this table are given for $B_z = B_0$ and $X = X_0$.  The saturation
level of the Maxwell and Reynolds stresses and the magnetic Reynolds
number are also listed in the table.  No data in the columns for the
saturation level means that it is less than $10^{-8}$.  The time
average is taken over the last 20 orbits for all models in this
section.

Figure \ref{fig:wm-t-s} shows the time evolution of the magnetic stress
for models S1 ($X_0 = 0$), S2 ($X_0 = 2$), and S3 ($X_0 = 4$).
All the other parameters for these models are identical
($\beta_0 = 3200$ and $Re_{M0} = 100$).
Because the magnetic Reynolds number for these models is very large,
the Hall effect dominates the evolution of the MRI.
The stress in the $X_0 = 4$ run has large amplitude time variations
and a higher saturation level than that in the $X_0 = 0$ run.
Thus, the Hall effect enhances the saturation level of the stress,
as in the uniform $B_z$ cases.
The difference in the time average $\langle \negthinspace \langle w_M
\rangle \negthinspace \rangle$ is a factor of 3 between the $X_0 = 0$
and 4 runs.
This saturation level is an order of magnitude lower than the uniform
$B_z$ models with the same $\beta_0$ and $Re_{M0}$.

During the linear phase, the MRI grows most rapidly in the half region with
positive $X$, because most of the region with $X < 0$ is linearly stable
for the MRI or has a smaller growth rate (see Paper I).
The amplified magnetic field gradually affects the structure of the
other ($X < 0$) half region, and after several orbits the entire region becomes
turbulent.
Figure \ref{fig:s3} shows images of the magnetic energy during the
turbulent phase for model S3 ($X_0 = 4$) at 35, 40, 45, and 50 orbits.
The magnetic field vectors in $x$-$z$ plane are also shown by arrows.
A large time variation in the stress occurs near 35 orbits (see
Fig. \ref{fig:wm-t-s}),
however from figure \ref{fig:s3} no channel flow is evident at this time.
Instead disorganized MHD turbulence is sustained throughout the evolution.
The amplitude of time variation in the stress is slightly smaller
compared with uniform $B_z$ cases.
At the turbulent phase, the initial distribution of the vertical field
disappears completely, and the positive and negative $X$ regions are
randomly distributed.
The effective Hall parameter at the nonlinear stage is very small and
negative for this case, $\langle \negthinspace \langle X_{\rm eff}
\rangle \negthinspace \rangle = - 0.018$. 
At 50 orbits, the spatial dispersion is large in the magnetic pressure
($\langle \delta P_{\rm mag}^2 \rangle^{1/2} / \langle P_{\rm mag}
\rangle = 1.0$) but small in the density and gas pressure ($\langle
\delta P^2 \rangle^{1/2} / \langle P \rangle = 0.12$ and $\langle \delta 
\rho^2 \rangle^{1/2} / \langle \rho \rangle = 0.069$).
These numbers are similar to the uniform $B_z$ cases during phases of
low magnetic energy.

Since there is no net-flux for this field configuration,
the magnetic field within the shearing box can completely die away.  In
fact, the magnetic energy during the nonlinear regime is decreasing in
time for models with small $Re_{M0}$.  Figure \ref{fig:em-t-s} shows
the time evolution of the magnetic energy for models with $Re_{M0} =
1$.  Ohmic dissipation affects the linear characters of the MRI in
these cases.  When the Hall parameter is small, models S8 ($X_0 = 0$)
and S9 ($X_0 = 2$), the magnetic energy is amplified by the MRI during
the linear phase, but this amplified field is not sustained.  After a
few tens of orbits, the magnetic energy and stress decrease due to
ohmic dissipation until the end of the calculation.  In the $X_0 = 4$
run (S10), however, the amplified magnetic energy is sustained for at
least 50 orbits even with $Re_{M0} = 1$.  The time evolution of run S10
is quite similar to the less resistive models shown in Figure
\ref{fig:wm-t-s}.  The saturation level of the Maxwell stress in model
S10 is $\langle \negthinspace \langle w_M \rangle \negthinspace \rangle
/ P_0 = 0.033$ and this is comparable to those in models S3 ($Re_{M0} =
100$) and S6 ($Re_{M0} = 10$).  
This enhancement is caused by a faster linear
growth rate for the MRI at larger $X_0$ (Paper I).  Once the field is
amplified, the efficiency of the Hall term and ohmic dissipation is
reduced, and thus turbulence can be sustained.

Table \ref{tbl:s2} lists the time- and volume-averaged quantities for
the models with $Re_{M0} = 100$ (S1 and S3) and $Re_{M0} = 1$
(S8 and S10).
Models S1 and S8 do not include the Hall effect ($X_0 = 0$), whereas
models S3 and S10 have $X_0 = 4$.
The properties of the turbulence in the $X_0 = 4$ runs (S3 and S10) are
nearly identical.
The saturation level of the magnetic energy is $\langle \negthinspace
\langle B^2 / 8 \pi \rangle \negthinspace \rangle / P_0 \approx 0.1$,
which is 3 times larger than the perturbed kinetic energy.
The ratio of the Maxwell and Reynolds stress is $\langle \negthinspace
\langle w_M \rangle \negthinspace \rangle / \langle \negthinspace
\langle w_R \rangle \negthinspace \rangle \approx 4$. 
The ratio of each component of the magnetic energy in the turbulence is
$\langle \negthinspace \langle B_x^2 \rangle \negthinspace \rangle :
\langle \negthinspace \langle B_y^2 \rangle \negthinspace \rangle :
\langle \negthinspace \langle B_z^2 \rangle \negthinspace \rangle \approx
2:18:1$, and this is quite similar to that in the uniform $B_z$ models.
Since the magnetic Reynolds number in the nonlinear regime $\langle
\negthinspace \langle Re_{M} \rangle \negthinspace \rangle$ is large,
ohmic dissipation is ineffective.  Moreover,
the Hall effect may also be unimportant in the nonlinear regime, because
the effective Hall parameter is small $\langle \negthinspace \langle
X_{\rm eff} \rangle \negthinspace \rangle \approx - 0.02$.
The saturated quantities in model S1 are also close to the $X_0 = 4$ runs.
For these three models, the saturated level of $\alpha$ is of the order of
0.01.

The turbulence in model S8, on the other hand, is strongly affected by
ohmic dissipation.  The stress $\alpha = 7.8 \times 10^{-6}$ at 50
orbits is much smaller than the other runs, and is decreasing in time.
The properties of the turbulence, such as $\langle \negthinspace
\langle B_x^2 \rangle \negthinspace \rangle / \langle \negthinspace
\langle B_z^2 \rangle \negthinspace \rangle$, are also very different
from the other runs.  The magnetic Reynolds number in the nonlinear
regime is still small $\langle \negthinspace \langle Re_{M} \rangle
\negthinspace \rangle = 1.2$, which means the ohmic dissipation is
important throughout the evolution of this model.

\subsection{Critical Magnetic Reynolds Number}

Here we estimate the critical value of the magnetic Reynolds number
$Re_{M0,{\rm crit}}$ for zero-net flux $B_z$ models.
Figure \ref{fig:wm-rm-s} shows the saturation level of the Maxwell
stress for the models without the Hall term ($X_0 = 0$).
Three different cases with $\beta_0 = 800$, 3200, and 12800 are shown in
this figure.
When $Re_{M0} = 100$, the stress for all the models is more than $0.01$, 
so that angular momentum transport is efficient.
If the magnetic Reynolds number is below a critical value, the stress
drops dramatically.
For the $\beta_0 = 3200$ runs, the stress is $\langle \negthinspace
\langle w_M \rangle \negthinspace \rangle / P_0 = 0.018$ and $2.0 \times
10^{-5}$ at $Re_{M0} = 10$ and 3, respectively.
Thus the stress decreases about 3 orders of magnitude with a small
difference in $Re_{M0}$.

The critical value is $Re_{M0,{\rm crit}} \sim 10$ for $\beta_0 = 3200$,
but this value is found to depend on the initial field strength.
The condition for $\langle \negthinspace \langle w_M \rangle
\negthinspace \rangle / P_0 > 0.01$, for example, is $Re_{M0} \ge 30$,
10, and 3 for $\beta_0 = 800$, 3200, and 12800, respectively.
When $Re_{M0} \ge Re_{M0,{\rm crit}}$, the saturation level of the
Maxwell stress is nearly independent of $\beta_0$ (Hawley, Gammie, \&
Balbus 1996) and this is different from the uniform $B_z$ case.
In order to initiate the turbulence, the MRI must grow faster than the
dissipation rate of the initial field.
Thus, the critical value for the instability should be given by
\begin{equation}
 \frac{v_{{\rm A}0} L}{\eta} = Re_{M0} \sqrt{\beta_0} 
\left(\frac{L}{L_x}\right) \sim const. ~,
\end{equation}
(where $L$ is the length over which the initial field varies) so that
the critical value of $Re_{M0}$ is proportional to
$\beta_0^{-1/2} L^{-1}$. 
This idea is roughly consistent with the results shown in Figure
\ref{fig:wm-rm-s}.
If $L$ decreases, the critical value should increase.
If fact, when we use a field distribution $B_z(x) = B_0 \sin (6 \pi x /
L_x)$ (i.e., $L = L_x/3$), the criterion becomes $Re_{M0} > 30$ for
$\beta_0 = 3200$.
This suggests that if the zero-net flux magnetic field has small-scale
structure initially, then the ohmic dissipation can suppress the
initiation of the MRI with a small magnetic diffusivity $\eta$.

Next we examine the effect of the Hall term on the critical magnetic
Reynolds number.
Figure \ref{fig:em-rm-s} shows the saturated magnetic energy as a
function of the initial magnetic Reynolds number $Re_{M0}$. 
For comparison, the magnetic Reynolds number $Re_{M0}'$ used in
Fleming et al. (2000) is also shown in the figure.
The initial field strength for all the models in this figure is $\beta_0 
= 3200$.
Open circles denotes the models without the Hall term.
For $X_0 = 0$, as discussed above, the saturation level is almost
constant for $Re_{M0} \ge 10$, but decreases as the magnetic Reynolds
number decreases if $Re_{M0} < 10$.
When $Re_{M0} = 0.3$, the magnetic energy does not show any increase
during the evolution, and almost dies out: $\langle \negthinspace \langle
B^2 / 8 \pi \rangle \negthinspace \rangle / P_0 \sim 10^{-12}$ at the
end of the calculation.
The critical value for significant turbulence, $\langle
\negthinspace \langle B^2 / 8 \pi \rangle \negthinspace \rangle / P_0 >
0.01$, is $Re_{M,{\rm crit}} \sim 10$.
This value corresponds to $Re'_{M0,{\rm crit}} \sim 3 \times 10^4$,
which is consistent with the results of Fleming et al. (2000).

For the $X_0 = 2$ runs depicted by filled triangles, the effect of the
Hall term is not large; the critical magnetic Reynolds number is
still $Re_{M0,{\rm crit}} \sim 10$.
However, if the size of the Hall term increases, we find that
$Re_{M,{\rm crit}}$ shifts to smaller values.
The $X_0 = 4$ runs are shown by filled circles in this figure.
The saturation level in the $Re_{M0} = 1$ run is increased to $\langle 
\negthinspace \langle B^2 / 8 \pi \rangle \negthinspace \rangle / P_0
= 0.088$.
But the $Re_{M0} = 0.3$ run shows no growth of the MRI.
Thus, the critical value for $\langle \negthinspace \langle B^2 / 8 \pi
\rangle \negthinspace \rangle / P_0 > 0.01$ becomes $Re_{M0,{\rm crit}}
\sim 1$, which is an order of magnitude smaller than the case without
the Hall term.

The Hall parameter $X_0$ can take a large range of values when
$Re_{M0}$ is small, because the maximum value is $X_{0} \sim
100/Re_{M0}$ in the Hall regime (Paper I).  The negative $X$ region ($x
> 0$) is linearly stable when $X_0 > 4$, however this is potentially
offset by the larger linear growth rate in the regions with a large
positive Hall parameter.  We have calculated models with $X_0 = 100$
(filled squares in Fig. \ref{fig:em-rm-s}) and 1000 (cross).  For
$Re_{M0} = 0.3$, the saturation level is enhanced by many orders of
magnitude when $X_0 = 100$, but the stress is still very small $\langle
\negthinspace \langle w_M \rangle \negthinspace \rangle / P_0 = 7.0
\times 10^{-5}$ compared with the less resistive models ($Re_{M0} \ge
10$).  For more resistive models ($Re_{M0} = 0.1$), the MRI does not
operate even with Hall parameters as large as $X_0 = 100$ and 1000.
Note that in the model with $Re_{M0} = 0.1$ and $X_0 = 1000$ (S16), the
critical wavelength is longer than the scale height of the disk,
meaning such large values of $X_0$ cannot result in enhanced transport
in real disks.  Therefore, the change in the critical magnetic Reynolds
number due to the Hall term is at most an order of magnitude for
zero-net flux vertical fields.

\section{SIMULATIONS WITH A UNIFORM TOROIDAL FIELD}

Lastly we investigate the evolution of the MRI starting with a uniform
toroidal field, $B_y = B_0$.
Table \ref{tbl:y} lists the models calculated with this field geometry.
Axisymmetric perturbations are stable if the field is purely toroidal. 
Since no linear dispersion relation has been obtained for the MRI
for nonaxisymmetric
perturbations including nonideal MHD effects, the characteristic length
of the MRI defined as $\lambda_{\rm MRI} \equiv 2 \pi v_{\rm A0} /
\Omega$ is listed in the table instead of $\lambda_{\rm crit}$ and
$\sigma_{\max}$.
The length $\lambda_{\rm MRI}$ corresponds to the wavelength in the
azimuthal direction of the most unstable mode.
The table gives the saturation level of the Maxwell stress $\langle
\negthinspace \langle w_M \rangle \negthinspace \rangle / P_0$, the
Reynolds stress $\langle \negthinspace \langle w_R \rangle \negthinspace
\rangle / P_0$, and the magnetic Reynolds number $\langle \negthinspace
\langle v_{{\rm A}z}^2 / \eta \Omega \rangle \negthinspace \rangle$, if
they are more than $10^{-8}$.
In some cases the uniform $B_y$ models must be evolved for very long times
to reach a saturated nonlinear regime.
Time averages are taken over the last 40 orbits for all the models in
this section.

We still use the Hall parameter $X_0$ defined by equation (\ref{eqn:x})
as a model parameter. 
Balbus \& Terquem (2001) examined the Hall effect on the nonaxisymmetric
behavior of linear perturbations.
The critical wavenumber for the MRI becomes longer or shorter
depending on the sign and size of the Hall parameter.  The effective
Hall parameter, $Ha$, for nonaxisymmetric disturbances with a
wavenumber {\boldmath $k$} is given by 
\begin{equation}
Ha \equiv \frac{c (\mbox{\boldmath $k$} \cdot \mbox{\boldmath $B$}) 
(\mbox{\boldmath $k$} \cdot \mbox{\boldmath $\Omega$})} 
{2 \pi e n_{e0} \Omega^2} = X_0 
\left( \frac{k_y v_{{\rm A}0}} {\Omega} \right)
\left( \frac{k_z v_{{\rm A}0}} {\Omega} \right) ~.
\end{equation}
Thus, the parameter $Ha$ depends on the wavenumber of each mode.
We begin the simulations with small random perturbations, so
that modes with every allowed {\boldmath $k$} are included initially.
We expect the most unstable mode to dominate
the linear phase of the MRI.
In the ideal MHD limit for a uniform $B_y$, the most unstable mode
has a characteristic wavenumber $k_y \sim v_{{\rm A}0} / \Omega$ in the
azimuthal direction, but no scale in the vertical direction $k_z
\rightarrow \infty$ (Balbus \& Hawley 1991).
In the numerical calculations, the grid resolution constrains the
minimum length in the vertical direction, which is typically $k_z v_{{\rm
A}0} / \Omega \sim 10$.
Because the parameter $Ha$ is larger than $X_0$ by the factor $k_z
v_{{\rm A}0} / \Omega$, we expect the Hall term will have an effect
even with small $X_0$.

Figure \ref{fig:wm-t-y} shows the time evolution of the Maxwell stress
for the models with $X_0 = 0$, 0.2, and 0.4, which are models Y2, Y3,
and Y4, respectively.  All the models have the same field strength and
the same magnetic Reynolds number, $\beta_0 = 100$ and $Re_{M0} =
100$.  The evolution is dominated by the Hall effect because ohmic
dissipation is inefficient ($Re_{M0} = 100 \gg 1$).  We find the
timescale to reach saturation is very sensitive to the Hall parameter.
Without the Hall term ($X_0 = 0$) it requires 15 orbits, but in the
$X_0 = 0.4$ run the turbulent state begins at about 8 orbits.  Thus, as
discussed above, the linear phase of the nonaxisymmetric MRI is
affected by even small $X_0$.

Despite the sensitivity of the linear growth rates to $X_0$, the
properties of the saturated turbulence are almost independent of
$X_0$.  The saturation level with $X_0 = 0.4$ is slightly higher than
$X_0 = 0$.  The amplitude of time variability is much
smaller than those in uniform vertical field cases, and the frequency is 
higher (see Fig.
\ref{fig:wm-t-z}).  Figure \ref{fig:y4} illustrates the evolution of
the magnetic energy in model Y4 ($\beta_0 = 100$, $Re_{M0} = 100$, and
$X_0 = 0.4$) taken along a slice of constant $y$.  The contours
show the azimuthal component of the magnetic energy with logarithmic
spacing, and the velocity field is shown by arrows.  The growth of
disturbances starts from modes with large {\boldmath $k$}, typically
$k_x \sim k_z \sim 5$ for this case (Fig. \ref{fig:y4}, {\it top-left
panel}).  At about 10 orbits the MRI saturates, and MHD turbulence
persists until the end of the simulation at 100 orbits.  No growth of
the two-channel flow is evident.  Small-scale growth of the MRI occurs
everywhere and this makes small amplitude and frequent time
variations.  At 20 orbits, the spatial dispersion in the magnetic
pressure ($\langle \delta P_{\rm mag}^2 \rangle^{1/2} / \langle P_{\rm
mag} \rangle = 1.3$) is larger than those in the density and gas
pressure ($\langle \delta P^2 \rangle^{1/2} / \langle P \rangle = 0.15$
and $\langle \delta \rho^2 \rangle^{1/2} / \langle \rho \rangle =
0.094$).  These numbers are almost constant throughout the evolution
and similar to the zero-net flux $B_z$ cases.

Ohmic dissipation is found to make a large difference in the saturation
level of the magnetic energy and stress in these pure azimuthal field
runs, as it did for models with other field configurations.  Because
the most unstable mode for nonaxisymmetric perturbation has a larger
$k_z$ than the axisymmetric mode on a vertical field, the effect of
ohmic dissipation is expected to be important for smaller $\eta$, or a
larger $Re_{M0}$.  Figure \ref{fig:bp-t-y} shows the time evolution of
the poloidal component of the magnetic energy $\langle (B_x^2 + B_z^2)
/ 8 \pi \rangle$ for models with different magnetic Reynolds number
$Re_{M0}$.  The magnetic field strength and the Hall parameter are
$\beta_0 = 100$ and $X_0 = 0.4$ for all the models in this figure.  The
initial magnetic field is purely toroidal $\langle B_y^2 / 8 \pi
\rangle / P_0 = 0.01$.  Initially, the poloidal field is generated by
small perturbations from the toroidal field, and thus much smaller than
the azimuthal component $\langle (B_x^2 + B_z^2) / 8 \pi \rangle / P_0
\sim 10^{-7}$.

When $Re_{M0} \ge 30$ the magnetic energy is amplified by the MRI in the 
linear phase, with
the linear growth starting earlier with larger $Re_{M0}$.
The saturation level of the poloidal field energy is comparable to the
initial toroidal field energy.
The toroidal field is also amplified by an order of magnitude, and
it dominates the other components.
The $Re_{M0} = 100$ and 30 runs evolve to the same saturation level,
and the Maxwell stress $\langle \negthinspace \langle w_M \rangle
\negthinspace \rangle / P_0$ is of the order of 0.01.
When $Re_{M0} = 10$, the amplified magnetic energy is sustained, but the 
saturation level of the poloidal field energy $\langle (B_x^2 + B_z^2) /
8 \pi \rangle / P_0$ is 3 orders of magnitude smaller.
The toroidal field is not amplified at all, so that the total magnetic
energy is unchanged from its initial value.
If the magnetic Reynolds number is less than 3, the initial
perturbations are decaying until the end of the calculation.

The time- and volume-averaged properties of the turbulence driven by the
MRI in the azimuthal field runs are listed in Table
\ref{tbl:y2}.
Models Y2 ($X_0 = 0$) and Y4 ($X_0 = 0.4$) show little effect from ohmic
dissipation ($Re_{M0} = 100$).
The more resistive ($Re_{M0} = 10$) models Y8 ($X_0 = 0$) and Y10 ($X_0
= 0.4$) are also listed in the table.
Quantities in models with the same $Re_{M0}$ are
similar, which suggests that the nonlinear effect of the Hall term is
small for uniform $B_y$ models.
In fact, the effective Hall parameter is small $\langle \negthinspace
\langle X_{\rm eff} \rangle \negthinspace \rangle = - 0.011$ and
$-0.0012$ for models Y4 and Y10.
For the $Re_{M0} = 100$ runs, the turbulence is significant and the
stress $\alpha$ is larger than 0.01.
The magnetic energy distribution $\langle \negthinspace \langle B_x^2
\rangle \negthinspace \rangle : \langle \negthinspace \langle B_y^2
\rangle \negthinspace \rangle : \langle \negthinspace \langle B_z^2
\rangle \negthinspace \rangle \approx 3 : 22 : 1$ and the perturbed
kinetic energy distribution $\langle \negthinspace \langle \rho v_x^2
\rangle \negthinspace \rangle : \langle \negthinspace \langle \rho
\delta v_y^2 \rangle \negthinspace \rangle : \langle \negthinspace
\langle \rho v_z^2 \rangle \negthinspace \rangle \approx 2 : 2 : 1$ are
similar to those in active turbulence of the other initial field
geometries. 
For the $Re_{M0} = 10$ runs, the poloidal field energy is much smaller
than the toroidal field energy and the ratio is $\langle \negthinspace
\langle B_y^2 \rangle \negthinspace \rangle / \langle \negthinspace
\langle B_z^2 \rangle \negthinspace \rangle \approx 3 \times 10^3$.
The kinetic energy is comparable to the poloidal field energy.
The stress $\alpha$ is about $10^{-5}$ so that the angular momentum
transport is inefficient for these models.

\subsection{Critical Magnetic Reynolds Number}

Here we consider the critical magnetic Reynolds number for a disk with 
a toroidal field.
Figure \ref{fig:wm-rm-y} shows the saturation level of the Maxwell
stress as a function of the initial magnetic Reynolds number $Re_{M0}$.
All the models in this figure are without the Hall term, only the
effect of ohmic dissipation is included.
The results with different field strength $\beta_0 = 100$, 400, and 1600
are shown.
The critical magnetic Reynolds number for significant stress (e.g.,
$\langle \negthinspace \langle w_M \rangle \negthinspace \rangle / P_0 >
0.01$) is $Re_{M0,{\rm crit}} \sim 30$, and this is
independent of the initial field strength $\beta_0$.
The saturation level is also independent of $\beta_0$, and typically
$\langle \negthinspace \langle w_M \rangle \negthinspace \rangle / P_0
\sim 0.04$.
For ideal MHD, the saturation level is close to this value
(Hawley et al. 1995).

If $Re_{M0}$ is less than 30, the evolution of the disk is quite
different.
Ohmic dissipation suppresses the MRI,
and the stress in the nonlinear regime is very small and of the order of
$10^{-6}$ -- $10^{-5}$.
The timescale to reach the saturation depends on both the initial field
strength and the magnetic Reynolds number.
When $Re_{M0} = 30$, for example, the saturation occurs at 25
orbits for the $\beta_0 = 100$ run (Y5) but it takes more than 200
orbits for the $\beta_0 = 1600$ run (Y24).

Next we consider the effect of the Hall term on the critical magnetic
Reynolds number.  Figure \ref{fig:bp-rm-y} shows the saturation level
of the poloidal magnetic energy $\langle \negthinspace \langle (B_x^2 +
B_z^2)/8 \pi \rangle \negthinspace \rangle / P_0$ for models including
the Hall effect.  The initial field strength is the same ($\beta_0 =
100$) for all the models.  Open circles depict the saturation level in
models without the Hall term.  When $Re_{M0} \ge 30$, the saturation
level is independent of $Re_{M0}$, and the poloidal field energy is of
the order of 0.01.  If $Re_{M0} < 30$, the turbulence in the nonlinear
regime is reduced by ohmic dissipation, and the saturation level drops
dramatically.  Filled triangles and circles denote the $X_0 = 0.2$ and
0.4 runs, respectively.  We find that the critical magnetic Reynolds
number is not affected by the size of the Hall parameter.  Even for
very large $X_0$ cases ($X_0 = 10$ and 100), the behavior at $Re_{M0}
\le 10$ is unchanged.  Therefore, the critical value for uniform $B_y$
models is $Re_{M0,{\rm crit}} \sim 30$ for any $\beta_0$ and $X_0$.
This is an order of magnitude larger than for uniform $B_z$ models.
For comparison, the magnetic Reynolds number $Re'_{M0}$ used in Fleming
et al. (2000) is also shown in this figure.  The critical value for
$Re'_{M0}$ is about $3 \times 10^3$, however this value should depend
on the initial field strength.

\section{DISCUSSION}

\subsection{Critical Magnetic Reynolds Number at the Nonlinear Stage}


The critical magnetic Reynolds number discussed in sections 3.4, 4.1,
and 5.1 is defined using the magnetic field strength in the initial
state.  However, the initial magnetic field in accretion disks is
highly uncertain, and only the field strength in the nonlinear,
saturated state of the instability may be observable.  We find that the
magnetic Reynolds number defined using the time- and volume-averaged
vertical magnetic field strength in the nonlinear regime, that is
$\langle \negthinspace \langle v_{{\rm A}z}^2 / \eta \Omega \rangle
\negthinspace \rangle$, characterizes the saturation amplitude of the
MRI very well.  Tables \ref{tbl:z} -- \ref{tbl:zl}, \ref{tbl:s}, and
\ref{tbl:y} list $\langle \negthinspace \langle v_{{\rm A}z}^2 / \eta
\Omega \rangle \negthinspace \rangle$ for all models calculated in this
paper.  Figure \ref{fig:alpha-rmz} shows the saturation level of the
stress $\alpha = \langle \negthinspace \langle w_{xy} \rangle
\negthinspace \rangle / P_0$ as a function of $\langle \negthinspace
\langle v_{{\rm A}z}^2 / \eta \Omega \rangle \negthinspace \rangle$ for
all the models.  Circles, triangles, and squares denote models started
with a uniform vertical, zero-net flux vertical, and a uniform toroidal
field, respectively.  Filled marks denote models that include the Hall
term.

The dependence of $\alpha$ on the magnetic Reynolds number is clearly
evident.  A stress larger than 0.01 requires $\langle \negthinspace
\langle v_{{\rm A}z}^2 / \eta \Omega \rangle \negthinspace \rangle >
1$.  The solid arrows denote the saturation levels obtained for ideal
MHD ($v_{\rm A}^2 / \eta \Omega \rightarrow \infty$), where the upper
and lower arrows ($\alpha = 0.29$ and 0.044) are the average of uniform
$B_z$ runs and uniform $B_y$ runs, respectively, taken from Hawley et
al. (1995).  The saturation level when $\langle \negthinspace \langle
v_{{\rm A}z}^2 / \eta \Omega \rangle \negthinspace \rangle > 1$ is
almost the same as the ideal MHD cases, and thus the stress is nearly
independent of the magnetic Reynolds number in this regime.  However,
when $\langle \negthinspace \langle v_{{\rm A}z}^2 / \eta \Omega
\rangle \negthinspace \rangle < 1$, the stress decreases as the
magnetic Reynolds number decreases.  If turbulence cannot be sustained
by the MRI, the magnetic energy and stress both decrease in time.  In
some models with $\langle \negthinspace \langle v_{{\rm A}z}^2 / \eta
\Omega \rangle \negthinspace \rangle < 1$, the magnetic energy is
decaying at the end of the calculation, and the system is evolving
toward the direction $\langle \negthinspace \langle w_{xy} \rangle
\negthinspace \rangle \propto \langle \negthinspace \langle v_{{\rm
A}z}^2 / \eta \Omega \rangle \negthinspace \rangle$ shown by dashed
arrow in the figure.

From Figure \ref{fig:alpha-rmz}, the saturation level of the stress is
approximately given by 
\begin{equation}
\alpha \approx \alpha_{\rm MRI} 
\min \left( 1 ~,~ \frac{v_{{\rm A}z}^2}{\eta \Omega} \right) ~,
\end{equation}
where $\alpha_{\rm MRI}$ is the stress in the ideal MHD limit ($v_{{\rm
A}z}^2 / \eta \Omega \gg 1$).  For a vertical field with zero-net flux
or a purely toroidal field, the dispersion in the saturation level is
quite small: from Tables \ref{tbl:s2} and \ref{tbl:y2} the averages of
the stress are nearly the same $\alpha_{\rm MRI} = 0.0395 \pm 0.0120$
and $0.0372 \pm 0.0106$ for these runs.  On the other hand, when the
disk has net-flux of $B_z$, the saturation levels are less uniform
because they depend on the vertical field strength, the vertical box
size (Hawley et al. 1995), and the Hall parameter.  The stress is
$\alpha_{\rm MRI} \sim 0.01 - 1$, and can be larger than zero-net
vertical flux models.  The stress obtained in Fleming et al. (2000) is
shown by crosses in this figure; our results are consistent with this
earlier work.

Figure \ref{fig:alpha-rmz} shows that the vertical field is a key
quantity for predicting the saturation amplitude of the MRI.
Nonaxisymmetric modes of the MRI are unstable when azimuthal field is
present, but because these modes have a large wavenumber in the
vertical direction, they are suppressed by a smaller resistivity than
axisymmetric modes.  Therefore the suppression of the axisymmetric
instability, which depends on the vertical field strength, determines
the critical magnetic Reynolds number.  This is true so long as the
azimuthal field gives an Alfv{\'e}n speed that is subthermal $v_{{\rm
A}y} < c_s$ (Blaes \& Balbus 1994), which is always true in our
calculations.  Thus, the effect of nonideal MHD on the saturation
amplitude of the MRI is well characterized by a magnetic Reynolds
number defined as $v_{{\rm A}z}^2 / \eta \Omega$, with a value larger
than unity required for turbulence and significant transport.

The critical value $\langle \negthinspace \langle v_{{\rm A}z}^2 /
\eta \Omega \rangle \negthinspace \rangle \sim 1$ is independent of the
initial field strength and geometry, and also the Hall parameter.
The toroidal component of the magnetic energy is dominant, and an
order of magnitude larger than the vertical component in the turbulence
driven by the MRI.
The ratio $\langle \negthinspace \langle B^2 \rangle
\negthinspace \rangle / \langle \negthinspace \langle B_z^2 \rangle
\negthinspace \rangle \sim 30$ when $\langle \negthinspace \langle
v_{{\rm A}z}^2 / \eta \Omega \rangle \negthinspace \rangle > 1$.
Thus the critical value can be written as $\langle \negthinspace \langle
v_{\rm A}^2 / \eta \Omega \rangle \negthinspace \rangle \sim 30$.
If the magnetic Reynolds number $v_{\rm A}^2 / \eta \Omega$
is larger than about 30, that means the MRI would be operating in the disk.

\subsection{Limit on the Parameter Range of Numerical Simulations}

The constraint on the time step in the numerical algorithm used here
(see Paper I) is inversely proportional to the initial Hall parameter
and the field strength.  Thus, when the saturation level of the
magnetic field is high, and when the Hall parameter is large, this
constraint is very severe.  Thus, we have computed only a few models
with large $X_0$ and large $Re_{M0}$.

In paper I, we estimated the ratio of the Hall parameter to the
magnetic Reynolds number in a weakly ionized gas composed by ions,
electrons, and neutrals.  When the Hall effect is dominant, the Hall
parameter takes on values $Re_{M0}^{-1} \lesssim X_0 \lesssim 100
Re_{M0}^{-1}$.  We have computed models with small $Re_{M0} (\sim 0.1)$
and with the maximum value of $X_0 (\sim 1000)$ which show inefficient
angular momentum transport in this case.  Thus, we conclude that
increasing the Hall parameter $X_0$ within the allowed range of values
will not change the lower limit of the critical $Re_{M0}$.  In fact,
with large $X_0$ (model S16) the critical wavelength of the MRI is
larger than the scale height of the disk, so that growth of the MRI
cannot be expected for such a situation.

For the $Re_{M0} > 1$ runs, we examined the effect of a Hall parameter
that is of the order of unity, the maximum value allowed when $Re_{M0}
\gtrsim 100$.  However, if $1 \lesssim Re_{M0} \lesssim 10$, the Hall
parameter $X_0$ can take any value between 10 -- 100.  We have found in
this work that a larger Hall parameter enhances the saturation level of
the stress, especially in the case of a uniform $B_z$.  But note that
even without the Hall effect, significant turbulence is sustained and
the stress $\alpha$ is more than 0.01 in this regime.  Thus, other effects
(such as stronger vertical fields) could be as important as the Hall effect
for enhancing the stress.

\subsection{Application to Protoplanetary Disks}

At the low temperatures expected in protoplanetary disks, thermal
ionization is inefficient, except for the innermost regions within
about $r \sim 0.1$ AU of the central star.  The dominant ionization
sources in this case are nonthermal processes, such as X-rays, cosmic
rays, and radioactive elements.  By definition, protoplanetary disks
contain dust grains that eventually will agglomerate into
planetesimals.  Because recombination process on the surface of grains
can be important, the number density and size distribution of dust
grains has a large effect on the ionization fraction of the gas.
Unfortunately, there are many uncertainties regarding dust grains in
protoplanetary disks, and the characteristics of the grains vary in
time due to evolutionary effects.  For example, grains may grow in size
through mutual collisions, while the abundance of grains may decrease
due to sedimentation toward the midplane.  Thus, the effect of dust
grains on the ionization state of the gas could be reduced in disks in
the late stages of evolution.  Therefore, it is important to
investigate the ionization state of the disk in situations both with
and without dust grains.

Assuming there are no dust grains in the disk, the distribution of the
Hall parameter and the magnetic Reynolds number were calculated at the
midplane of some disk models in Paper I.  Inside $r_{H} \sim 80$ AU,
the Hall parameter is $|X| > 1$ and the Hall effect is the dominant
nonideal MHD effect when the field strength is $c_s / v_{\rm
A} = 10$ in the disk.  Ohmic dissipation is more efficient in higher
density (inner) regions.  The magnetic Reynolds number is less than
unity at $r < r_{O} \sim 6$ AU.  The critical radius $r_{O}$ depends on
the field strength in the disk, e.g., $r_{O} \sim 2$ AU and $r_{H} \sim
12$ AU for $c_s / v_{\rm A} = 1$.  In the outer disk $r > r_{H}$, both
the Hall effect and ohmic dissipation are unimportant, and thus the
disk is unstable to the MRI.

According to the results of the nonlinear simulations presented in this
paper, angular momentum transport is inefficient ($\alpha \ll
0.01$) when $Re_{M0} \lesssim 1$ for any initial field strength
$\beta_0$ and for any size of the  Hall parameter $X_0$.  Therefore, at
the region 0.1 AU $\lesssim r \lesssim 6$ AU, the MRI is suppressed by
ohmic dissipation and this region probably forms a ``dead zone''
(Gammie 1996).  The critical magnetic Reynolds number $Re_{M0,{\rm
crit}}$ for the onset of active turbulence is 1 -- 30 depending on the
strength and geometry of the magnetic field.  However, because $Re_{M}$
is a steeply decreasing function of $r$, the corresponding uncertainty
in the critical radius $r_{O}$ is at most a factor of 3 (see Fig. 2 in
Paper I).

In the region $r_{O} < r < r_{H}$, the Hall parameter is larger than
unity, and thus the Hall term can affect the evolution of the MRI.
Moreover ohmic dissipation is too small to suppress the MRI.  Although
the nonlinear behavior of the MRI depends on the field geometry, the
MRI will operate for most of the cases in this region.  If there is no
net flux in the vertical field (zero-net flux $B_z$ or uniform $B_y$
models), MHD turbulence is initiated by the MRI and sustained for the
values of $Re_{M0}$ and $X_0$ in this region.  Although this region is
unstable even without the Hall term, the stress could be enhanced due
to the Hall effect.  If the disk is threaded by a uniform vertical
field oriented in the opposite direction to the angular velocity vector
{\boldmath $\Omega$}, i.e., $X_0 \ll -1$, the linear growth of the MRI
is suppressed by the Hall effect, forming a ``dead zone" with no
angular momentum transport via magnetic stress.  Since suppression of
the MRI requires the Hall parameter is $X_0 < -4$ everywhere, this may
be difficult in actual accretion disks.  If the vertical field is
oriented in the same sense as {\boldmath $\Omega$}, even in only a
small part of the disk, the MRI can grow from this region.  The
unstable region could spread wider and eventually fill the entire
region, as demonstrated in the zero-net flux $B_z$ simulations.

In summary, we expect the outer regions of protoplanetary disks $ r >
r_{O}$ to be unstable to the MRI, with angular momentum transported
effectively by Maxwell stress.  Inside of $r_{O}$ is a dead zone unless
the temperature is $T \gtrsim 10^3$ K.  The critical radius $r_{O}$ is
a few AU and this is determined mainly by the ohmic dissipation.  The
typical size of protoplanetary disks is about 100 AU.  Most of
the disk is unstable if dust grains are not present due to settling,
or are too large to affect the ionization fraction.  The critical
radius $r_O$, which determines where the disk is unstable to the MRI
and therefore turbulent, depends strongly on the size distribution and
abundance of dust grains (Sano et al. 2000).   The  extent to which
grains are mixed vertically through the disk in turn depends strongly
on whether the gas is turbulent.  Hence the evolution of gas and dust
grains must be solved simultaneously in order to make a consistent
scenario of grain settling, growth, and ultimately planet formation.

\section{SUMMARY}

The Hall effect on the nonlinear evolution of the MRI has been
investigated using local 3D nonideal MHD simulations.  Various models
with different field strengths $\beta_0$, magnetic Reynolds number
$Re_{M0} = v_{{\rm A}0}^2 / \eta \Omega$, and Hall parameter $X_0$ have
been computed.  Our findings are summarized as follows.

\begin{enumerate}
\item
For uniform $B_z$ models, a positive (negative) $X_0$ enhances
(suppresses) the nonlinear turbulence in the disk generated by the MRI.
When $X_0 \ge 0$, the nonlinear evolution shows recurrent appearances
of the channel flow, and the saturated Maxwell stress is larger than that
in the negative $X_0$ case.
The saturation level of the magnetic energy and stress decreases
dramatically when $Re_{M0} < Re_{M0,{\rm crit}} \sim 1$.
This critical value is independent of both the field strength and the
size of the Hall term.
\item
For zero-net flux $B_z$ models, disorganized MHD turbulence is sustained 
in the nonlinear regime without the growth of the channel flow.
The critical magnetic Reynolds number $Re_{M0,{\rm crit}}$ for the
initial state depends on the initial field strength $\beta_0$ and the
Hall parameter $X_0$, and is in the range of 1 -- 30 for our models.
The Hall effect can reduce the critical value $Re_{M0,{\rm crit}}$,
however the difference is at most an order of magnitude.
\item
For uniform $B_y$ models, the effect of the Hall term can be seen only
during the linear growth phase of the instability.
Disorganized turbulence lasts more than hundred orbits when the 
initial magnetic Reynolds number is $Re_{M0} \gtrsim 30$, and this
critical value is independent of both the field strength and the size of
the Hall term.
The characteristics of the saturated turbulence are quite similar to those
in the zero-net flux $B_z$ models.
\item
The condition for turbulence and significant transport in the nonlinear regime
is found to be given by
$\langle \negthinspace \langle v_{{\rm A}z}^2 / \eta \Omega \rangle
\negthinspace \rangle \gtrsim 1$ (or $\langle \negthinspace \langle
v_{{\rm A}}^2 / \eta \Omega \rangle \negthinspace \rangle \gtrsim 30$),
where $v_{{\rm A}z}$ is the Alfv\'{e}n speed computed from only the vertical
field component in the nonlinear regime.
This is independent of the strength and geometry of the initial 
magnetic field and the Hall parameter. 
If the magnetic field strength in a disk is estimated observationally
and the magnetic Reynolds number $v_{\rm A}^2 / \eta \Omega$ is larger
than about 30, this would imply the MRI is operating in the disk.
\end{enumerate}
We have applied the results of our simulations to the dynamics of
protoplanetary disks.  We conclude the stability of such disks is
determined mainly by the distribution of the magnetic Reynolds number.
The Hall effect makes little change in the critical value of $Re_{M}$.
The size of the dead zone where the MRI may be suppressed by ohmic
dissipation is sensitive to the characteristics of dust grains in the
disk.  When small dust grains are well-mixed vertically in the disk,
the dead zone extends to a few tens of AU from the central star.
However, the dead zone is considerably smaller if small grains are not
present.

\acknowledgements
Computations were carried out on VPP300/16R and VPP5000 at the National
Astronomical Observatory of Japan and VPP700 at the Subaru Telescope,
NAOJ.  This work is supported by a grant from the NASA OSS program.

\clearpage 

\clearpage 

\begin{deluxetable}{ccccccclll}
\small
\tablecaption{Uniform $B_z$ Simulations (Standard Box Size and 
Resolution\tablenotemark{a} )}
\tablehead{
\colhead{Model} & 
\colhead{$\beta_0$} & 
\colhead{$Re_{M0}$} & \colhead{$X_0$} & 
\colhead{$\lambda_{\rm crit}/H$} & 
\colhead{$\sigma_{\max}/\Omega$} & 
\colhead{Orbits} &
\colhead{$\langle\negthinspace\langle w_M 
\rangle\negthinspace\rangle/P_0$} & 
\colhead{$\langle\negthinspace\langle w_R
\rangle\negthinspace\rangle/P_0$} & 
\colhead{$\langle\negthinspace\langle v_{{\rm A}z}^2 / \eta \Omega
\rangle\negthinspace\rangle$} 
}
\startdata
Z1 & 3200 & 100 & 4 & 0.14 & 0.75 & 25 & 
0.181 & 0.0348 & $5.92 \times 10^{3}$ \nl
Z2 & 3200 & 100 & 2 & 0.11 & 0.75 & 50 &
0.162 & 0.0338 & $4.41 \times 10^{3}$ \nl
Z3 & 3200 & 100 & 0 & 0.064 & 0.74 & 50 & 
0.108 & 0.0218 & $2.84 \times 10^{3}$ \nl
Z4 & 3200 & 100 & $-2$ & 0.0 & 0.70 & 50 &
0.0394 & 0.0124 & $1.23 \times 10^{3}$ \nl
Z5 & 3200 & 10 & 0 & 0.064 & 0.70 & 25 &
0.0896 & 0.0227 & 271 \nl
Z6 & 3200 & 1 & 4 & 0.15 & 0.57 & 25 &
0.184 & 0.0268 & 53.0 \nl
Z7 & 3200 & 1 & 2 & 0.12 & 0.51 & 25 &
0.116 & 0.0237 & 32.3 \nl
Z8 & 3200 & 1 & 0 & 0.091 & 0.43 & 25 &
0.0564 & 0.0140 & 15.6 \nl
Z9 & 3200 & 1 & $-2$ & 0.064 & 0.28 & 25 &
0.0272 & 0.00956 & 7.72 \nl
Z10 & 3200 & 0.3 & 0 & 0.22 & 0.20 & 50 &
0.0187 & 0.00426 & 1.48 \nl
Z11 & 3200 & 0.1 & 4 & 0.48 & 0.14 & 100 &
$5.05 \times 10^{-4}$ & 0.00179 & 0.138 \nl
Z12 & 3200 & 0.1 & 2 & 0.54 & 0.11 & 100 &
$1.16 \times 10^{-4}$ & $3.96 \times 10^{-4}$ & 0.108 \nl
Z13 & 3200 & 0.1 & 0 & 0.64 & 0.074 & 100 &
$3.07 \times 10^{-4}$ & 0.00136 & 0.122 \nl
Z14 & 3200 & 0.1 & $-2$ & 0.91 & 0.037 & 200 &
$6.44 \times 10^{-6}$ & $1.74 \times 10^{-5}$ & 0.101 \nl
Z15 & 800 & 1 & 2 & 0.25 & 0.51 & 25 &
0.892 & 0.0927 & 38.7 \nl
Z16 & 800 & 1 & 0 & 0.18 & 0.43 & 25 &
0.178 & 0.0322 & 9.00 \nl
Z17 & 800 & 1 & $-2$ & 0.13 & 0.28 & 25 &
0.0323 & 0.0117 & 3.12 \nl
Z18 & 12800 & 100 & 0 & 0.032 & 0.74 & 25 &
0.0251 & 0.00734 & $2.54 \times 10^{3}$ \nl
Z19 & 12800 & 10 & 0 & 0.032 & 0.70 & 25 &
0.0340 & 0.00944 & 375 \nl
Z20 & 12800 & 1 & 2 & 0.061 & 0.51 & 25 &
0.0350 & 0.00924 & 39.4 \nl
Z21 & 12800 & 1 & 0 & 0.045 & 0.43 & 25 &
0.0262 & 0.00734 & 27.9 \nl
Z22 & 12800 & 1 & $-2$ & 0.032 & 0.28 & 25 &
0.0133 & 0.00399 & 12.1 \nl
Z23 & 12800 & 0.3 & 0 & 0.11 & 0.20 & 50 &
0.0147 & 0.00391 & 3.86 \nl
Z24 & 12800 & 0.1 & 0 & 0.32 & 0.074 & 100 &
$4.09 \times 10^{-4}$ & $5.89 \times 10^{-5}$ & 0.127 \nl
\enddata
\tablenotetext{a}{Box size is $H \times 4H \times H$ and grid
 resolutions is $32 \times 128 \times 32$.}
\label{tbl:z}
\end{deluxetable}

\begin{deluxetable}{ccccccclll}
\small
\tablecaption{Uniform $B_z$ Simulations (High Resolution\tablenotemark{a} )}
\tablehead{
\colhead{Model} & 
\colhead{$\beta_0$} & 
\colhead{$Re_{M0}$} & \colhead{$X_0$} & 
\colhead{$\lambda_{\rm crit}/H$} & 
\colhead{$\sigma_{\max}/\Omega$} & 
\colhead{Orbits} &
\colhead{$\langle\negthinspace\langle w_M 
\rangle\negthinspace\rangle/P_0$} & 
\colhead{$\langle\negthinspace\langle w_R
\rangle\negthinspace\rangle/P_0$} & 
\colhead{$\langle\negthinspace\langle v_{{\rm A}z}^2 / \eta \Omega
\rangle\negthinspace\rangle$} 
}
\startdata
Z2H & 3200 & 100 & 2 & 0.11 & 0.75 & 10 &
0.128 & 0.0262 & $5.36 \times 10^{3}$ \nl
Z4H & 3200 & 100 & $-2$ & 0.0 & 0.70 & 10 &
0.0301 & 0.0102 & $1.35 \times 10^{3}$ \nl
\enddata
\label{tbl:zh}
\tablenotetext{a}{Box size is $H \times 4H \times H$ and grid
 resolutions is $64 \times 256 \times 64$.}
\end{deluxetable}

\begin{deluxetable}{ccccccclll}
\small
\tablecaption{Uniform $B_z$ Simulations (Large Box Size\tablenotemark{a} )}
\tablehead{
\colhead{Model} & 
\colhead{$\beta_0$} & 
\colhead{$Re_{M0}$} & \colhead{$X_0$} & 
\colhead{$\lambda_{\rm crit}/H$} & 
\colhead{$\sigma_{\max}/\Omega$} & 
\colhead{Orbits} &
\colhead{$\langle\negthinspace\langle w_M 
\rangle\negthinspace\rangle/P_0$} & 
\colhead{$\langle\negthinspace\langle w_R
\rangle\negthinspace\rangle/P_0$} & 
\colhead{$\langle\negthinspace\langle v_{{\rm A}z}^2 / \eta \Omega
\rangle\negthinspace\rangle$} 
}
\startdata
Z2L & 3200 & 100 & 2 & 0.11 & 0.75 & 25 &
0.159 & 0.0428 & $6.91 \times 10^{3}$ \nl
Z4L & 3200 & 100 & $-2$ & 0.0 & 0.70 & 25 &
0.0676 & 0.0215 & $2.47 \times 10^{3}$ \nl
\enddata
\label{tbl:zl}
\tablenotetext{a}{Box size is $2H \times 8H \times 2H$ and grid
 resolutions is $64 \times 256 \times 64$.}
\end{deluxetable}

\begin{deluxetable}{cllllll}
\small
\tablewidth{0pc} 
\tablecaption{Time- and Volume-Averaged Values in Uniform $B_z$
 Simulations}
\tablehead{
\colhead{Quantity} & 
\colhead{Z2} & \colhead{Z3} & \colhead{Z4} & 
\colhead{Z12} & \colhead{Z13} & \colhead{Z14}
}
\startdata
$\beta_0$ & 3200 & 3200 & 3200 & 3200 & 3200 & 3200 \nl
$Re_{M0}$ & 100 & 100 & 100 & 0.1 & 0.1 & 0.1 \nl
$X_{0}$ & 2 & 0 & $-2$ & 2 & 0 & $-2$ \nl
$\langle\negthinspace\langle B_x^2 / 8 \pi 
\rangle\negthinspace\rangle / P_0$ &  
0.0435 & 0.0272 & 0.00974 & 
$3.79 \times 10^{-5}$ & $1.16 \times 10^{-4}$ & $1.89 \times 10^{-6}$ \nl
$\langle\negthinspace\langle B_y^2 / 8 \pi 
\rangle\negthinspace\rangle / P_0$ &  
0.314 & 0.210 & 0.0742 & 
$2.21 \times 10^{-4}$ & $6.64 \times 10^{-4}$ & $2.22 \times 10^{-5}$ \nl
$\langle\negthinspace\langle B_z^2 / 8 \pi 
\rangle\negthinspace\rangle / P_0$ &  
0.0139 & 0.00913 & 0.00386 & 
$3.37 \times 10^{-4}$ & $3.81 \times 10^{-4}$ & $3.14 \times 10^{-4}$ \nl
$\langle\negthinspace\langle \rho v_x^2 / 2
\rangle\negthinspace\rangle / P_0$ &  
0.0513 & 0.0333 & 0.0193 &
0.00154 & 0.00335 & $5.09 \times 10^{-5}$ \nl
$\langle\negthinspace\langle \rho \delta v_y^2 / 2
\rangle\negthinspace\rangle / P_0$ &  
0.0458 & 0.0301 & 0.0125 & 0.00129 & 0.00291 & $2.19 \times 10^{-5}$ \nl
$\langle\negthinspace\langle \rho v_z^2 / 2
\rangle\negthinspace\rangle / P_0$ &  
0.0160 & 0.0135 & 0.00816 & 0.00833 & 0.00812 & 0.00514 \nl
$\langle\negthinspace\langle P
\rangle\negthinspace\rangle / P_0$ &  
2.78 & 4.63  & 2.68 & 26.6 & 84.7 & 1.52 \nl
$\langle\negthinspace\langle w_M
\rangle\negthinspace\rangle / 
\langle\negthinspace\langle w_R
\rangle\negthinspace\rangle $ &  
4.78 & 4.95 & 3.19 & 0.294 & 0.226 & 0.371 \nl
$\langle\negthinspace\langle w_M
\rangle\negthinspace\rangle / 
\langle\negthinspace\langle B^2 / 8 \pi
\rangle\negthinspace\rangle $ &  
0.454 & 0.454 & 0.470 & 0.457 & 0.405 & 0.297 \nl
$\langle\negthinspace\langle P
\rangle\negthinspace\rangle / 
\langle\negthinspace\langle B^2 / 8 \pi
\rangle\negthinspace\rangle $ &  
7.81 & 19.5 & 32.0 & 
$1.04 \times 10^{5}$ & $1.12 \times 10^{5}$ & $7.03 \times 10^{4}$ \nl
$\langle\negthinspace\langle \rho \delta v^2 / 2
\rangle\negthinspace\rangle / 
\langle\negthinspace\langle B^2 / 8 \pi
\rangle\negthinspace\rangle $ &  
0.318 & 0.324 & 0.477 & 43.8 & 19.0 & 241 \nl
$\langle\negthinspace\langle B_x^2
\rangle\negthinspace\rangle / 
\langle\negthinspace\langle B_z^2
\rangle\negthinspace\rangle $ &  
3.12 & 2.98 & 2.52 & 0.112 & 0.304 & 0.00600 \nl
$\langle\negthinspace\langle B_y^2
\rangle\negthinspace\rangle / 
\langle\negthinspace\langle B_z^2
\rangle\negthinspace\rangle $ &  
22.5 & 23.0 & 19.2 & 0.657 & 1.74 & 0.0707 \nl
$\langle\negthinspace\langle v_x^2
\rangle\negthinspace\rangle / 
\langle\negthinspace\langle v_z^2
\rangle\negthinspace\rangle $ &  
3.21 & 2.46 & 2.36 & 0.184 & 0.413 & 0.00992 \nl
$\langle\negthinspace\langle \delta v_y^2
\rangle\negthinspace\rangle / 
\langle\negthinspace\langle v_z^2
\rangle\negthinspace\rangle $ &  
2.87 & 2.23 & 1.54 & 0.155 & 0.359 & 0.00426 \nl
$\langle\negthinspace\langle \beta
\rangle\negthinspace\rangle$ &  
104 & 316 & 448 & 
$6.78 \times 10^{4}$ & $1.71 \times 10^{5}$ & $4.55 \times 10^{3}$ \nl
$\langle\negthinspace\langle Re_M
\rangle\negthinspace\rangle$ &  
$1.27 \times 10^{5}$ & $8.02 \times 10^{4}$ & $2.88 \times 10^{4}$ & 
0.191 & 0.371 & 0.108 \nl
$\langle\negthinspace\langle X_{\rm eff}
\rangle\negthinspace\rangle$ &  
0.00277 & 0 & $-0.0344$ & 
1.41 & 0 & $-1.86$ \nl
$\alpha$ &  
0.195 & 0.130 & 0.0518 & 
$5.12 \times 10^{-4}$ & 0.00167 & $2.38 \times 10^{-5}$ \nl
\enddata
\label{tbl:z2}
\end{deluxetable}

\begin{deluxetable}{ccccccclll}
\small
\tablecaption{Zero-Net Flux $B_z$ Simulations (Standard Box Size and 
Resolution\tablenotemark{a} )}
\tablehead{
\colhead{Model} & 
\colhead{$\beta_0$} & 
\colhead{$Re_{M0}$} & \colhead{$X_0$} & 
\colhead{$\lambda_{\rm crit}/H$} & 
\colhead{$\sigma_{\max}/\Omega$} & 
\colhead{Orbits} &
\colhead{$\langle\negthinspace\langle w_M 
\rangle\negthinspace\rangle/P_0$} & 
\colhead{$\langle\negthinspace\langle w_R
\rangle\negthinspace\rangle/P_0$} & 
\colhead{$\langle\negthinspace\langle v_{{\rm A}z}^2 / \eta \Omega
\rangle\negthinspace\rangle$} 
}
\startdata
S1 & 3200 & 100 & 0 & 0.064 & 0.74 & 50 & 
0.0168 & 0.0130 & 630 \nl
S2 & 3200 & 100 & 2 & 0.11 & 0.75 & 50 &
0.0215 & 0.00673 & 647 \nl
S3 & 3200 & 100 & 4 & 0.14 & 0.75 & 50 & 
0.0520 & 0.0133 & $1.92 \times 10^{3}$ \nl
S4 & 3200 & 10 & 0 & 0.064 & 0.70 & 50 &
0.0177 & 0.00837 & 77.4 \nl
S5 & 3200 & 10 & 2 & 0.11 & 0.72 & 50 &
0.0244 & 0.00677 & 68.3 \nl
S6 & 3200 & 10 & 4 & 0.14 & 0.73 & 50 &
0.0427 & 0.0118 & 166 \nl
S7 & 3200 & 3 & 0 & 0.068 & 0.60 & 50 &
$2.01 \times 10^{-5}$ & $2.28 \times 10^{-6}$ & 0.00401 \nl
S8 & 3200 & 1 & 0 & 0.091 & 0.43 & 50 &
$3.95 \times 10^{-6}$ & $3.81 \times 10^{-6}$ & $2.63 \times 10^{-4}$ \nl
S9 & 3200 & 1 & 2 & 0.12 & 0.51 & 50 &
$8.83 \times 10^{-7}$ & $1.03 \times 10^{-6}$ & $6.66 \times 10^{-6}$ \nl
S10 & 3200 & 1 & 4 & 0.15 & 0.57 & 50 &
0.0326 & 0.00881 & 13.4 \nl
S11 & 3200 & 0.3 & 0 & 0.22 & 0.20 & 50 &
\nodata & \nodata & \nodata \nl
S12 & 3200 & 0.3 & 2 & 0.21 & 0.28 & 50 &
\nodata & \nodata & \nodata \nl
S13 & 3200 & 0.3 & 4 & 0.21 & 0.34 & 50 &
\nodata & \nodata & \nodata \nl
S14 & 3200 & 0.3 & 100 & 0.65 & 0.70 & 50 &
$7.02 \times 10^{-5}$ & $1.86 \times 10^{-7}$ & 0.00629 \nl
S15 & 3200 & 0.1 & 100 & 0.66 & 0.62 & 50 &
\nodata & \nodata & \nodata \nl
S16 & 3200 & 0.1 & 1000 & 2.0 & 0.74 & 50 &
\nodata & \nodata & \nodata \nl
S17 & 800 & 100 & 0 & 0.13 & 0.74 & 50 &
0.0173 & 0.0106 & 194 \nl
S18 & 800 & 30 & 0 & 0.13 & 0.73 & 50 &
0.0159 & 0.0130 & 59.8 \nl
S19 & 800 & 10 & 0 & 0.13 & 0.70 & 50 &
$1.41 \times 10^{-5}$ & $2.44 \times 10^{-6}$ & $7.37 \times 10^{-4}$ \nl
S20 & 800 & 3 & 0 & 0.14 & 0.60 & 50 &
$1.74 \times 10^{-7}$ & $3.29 \times 10^{-6}$ & $1.75 \times 10^{-5}$ \nl
S21 & 12800 & 100 & 0 & 0.032 & 0.74 & 100 &
0.0265 & 0.0134 & $4.31 \times 10^3$ \nl
S22 & 12800 & 10 & 0 & 0.032 & 0.70 & 100 &
0.0239 & 0.0152 & 358 \nl
S23 & 12800 & 3 & 0 & 0.034 & 0.60 & 100 &
0.0349 & 0.0153 & 160 \nl
S24 & 12800 & 1 & 0 & 0.045 & 0.43 & 100 &
$9.90 \times 10^{-6}$ & $4.15 \times 10^{-6}$ & 0.00179 \nl
S25 & 12800 & 0.3 & 0 & 0.11 & 0.20 & 100 &
\nodata & $1.27 \times 10^{-6}$ & $1.66 \times 10^{-6}$ \nl
\enddata
\tablenotetext{a}{Box size is $H \times 4H \times H$ and grid
 resolutions is $32 \times 128 \times 32$.}
\label{tbl:s}
\end{deluxetable}

\begin{deluxetable}{cllllc}
\small
\tablewidth{0pc} 
\tablecaption{Time- and Volume-Averaged Values in Zero-Net Flux $B_z$
 Simulations}
\tablehead{
\colhead{Quantity} & 
\colhead{S1} & \colhead{S3} & \colhead{S8} & 
\colhead{S10} & \colhead{Average\tablenotemark{a}}
}
\startdata
$\beta_0$ & 3200 & 3200 & 3200 & 3200 & \nodata \nl
$Re_{M0}$ & 100 & 100 & 1 & 1 & \nodata \nl
$X_{0}$ & 0 & 4 & 0 & 4 & \nodata \nl
$\langle\negthinspace\langle B_x^2 / 8 \pi 
\rangle\negthinspace\rangle / P_0$ &  
0.00584 & 0.0138 & $1.11 \times 10^{-8}$ & 0.00938 & 
$0.00821 \pm 0.00274 $\nl
$\langle\negthinspace\langle B_y^2 / 8 \pi 
\rangle\negthinspace\rangle / P_0$ &  
0.0404 & 0.115 & $3.71 \times 10^{-4}$ & 0.0743 & 
$0.0604 \pm 0.0237$ \nl
$\langle\negthinspace\langle B_z^2 / 8 \pi 
\rangle\negthinspace\rangle / P_0$ &  
0.00207 & 0.00605 & $8.24 \times 10^{-8}$ & 0.00416 &
$0.00339 \pm 0.00129$ \nl
$\langle\negthinspace\langle \rho v_x^2 / 2
\rangle\negthinspace\rangle / P_0$ &  
0.0281 & 0.0206 & $1.28 \times 10^{-4}$ & 0.0146 &
$0.0266 \pm 0.0119$ \nl
$\langle\negthinspace\langle \rho \delta v_y^2 / 2
\rangle\negthinspace\rangle / P_0$ &  
0.0199 & 0.0143 & $9.69 \times 10^{-7}$ & 0.0102 &
$0.0176 \pm 0.0075$ \nl
$\langle\negthinspace\langle \rho v_z^2 / 2
\rangle\negthinspace\rangle / P_0$ &  
0.00686 & 0.00766 & $7.95 \times 10^{-7}$ & 0.00565 &
$0.00824 \pm 0.00304$ \nl
$\langle\negthinspace\langle P
\rangle\negthinspace\rangle / P_0$ &  
2.58 & 1.94 & 1.04 & 3.32 & 
\nodata \nl
$\langle\negthinspace\langle w_M
\rangle\negthinspace\rangle / 
\langle\negthinspace\langle w_R
\rangle\negthinspace\rangle $ &  
1.29 & 3.91 & 1.04 & 3.70 & 
$2.51 \pm 0.98$ \nl
$\langle\negthinspace\langle w_M
\rangle\negthinspace\rangle / 
\langle\negthinspace\langle B^2 / 8 \pi
\rangle\negthinspace\rangle $ &  
0.362 & 0.403 & 0.0106 & 0.390 & 
$0.394 \pm 0.018$ \nl
$\langle\negthinspace\langle P
\rangle\negthinspace\rangle / 
\langle\negthinspace\langle B^2 / 8 \pi
\rangle\negthinspace\rangle $ &  
55.5 & 15.0 & $2.81 \times 10^3$ & 39.8 &
\nodata \nl
$\langle\negthinspace\langle \rho \delta v^2 / 2
\rangle\negthinspace\rangle / 
\langle\negthinspace\langle B^2 / 8 \pi
\rangle\negthinspace\rangle $ &  
1.18 & 0.330 & 0.349 & 0.364 & 
$0.869 \pm 0.447$ \nl
$\langle\negthinspace\langle B_x^2
\rangle\negthinspace\rangle / 
\langle\negthinspace\langle B_z^2
\rangle\negthinspace\rangle $ &  
2.83 & 2.28 & 0.134 & 2.26 &
$2.46 \pm 0.22$ \nl
$\langle\negthinspace\langle B_y^2
\rangle\negthinspace\rangle / 
\langle\negthinspace\langle B_z^2
\rangle\negthinspace\rangle $ &  
19.6 & 19.0 & $4.51 \times 10^3$ & 17.9 & 
$18.0 \pm 3.1$ \nl
$\langle\negthinspace\langle v_x^2
\rangle\negthinspace\rangle / 
\langle\negthinspace\langle v_z^2
\rangle\negthinspace\rangle $ &  
4.10 & 2.69 & 161 & 2.58 &
$3.15 \pm 0.46$ \nl
$\langle\negthinspace\langle \delta v_y^2
\rangle\negthinspace\rangle / 
\langle\negthinspace\langle v_z^2
\rangle\negthinspace\rangle $ &  
2.90 & 1.86 & 1.22 & 1.80 &
$2.10 \pm 0.32$ \nl
$\langle\negthinspace\langle \beta
\rangle\negthinspace\rangle$ &  
$1.23 \times 10^3$ & 204 & $2.81 \times 10^5$ & 562 &
\nodata \nl
$\langle\negthinspace\langle Re_M
\rangle\negthinspace\rangle$ &  
$1.52 \times 10^4$ & $4.51 \times 10^4$ & 1.19 & 288 & \nodata \nl
$\langle\negthinspace\langle X_{\rm eff}
\rangle\negthinspace\rangle$ &  
0 & $-0.0181$ & 0 & $-0.0335$ & \nodata \nl
$\alpha$ &  
0.0298 & 0.0653 & $7.76 \times 10^{-6}$ & 0.0414 &
$0.0385 \pm 0.0120$ \nl
\enddata
\tablenotetext{a}{Models with $\langle \negthinspace \langle v_{{\rm A}z}^2 
/ \eta \Omega \rangle \negthinspace \rangle > 1$ are considered for 
the average (S1 -- S6, S10, S17, S18, and  S21 -- S23).}
\label{tbl:s2}
\end{deluxetable}

\begin{deluxetable}{cccccclll}
\small
\tablecaption{Uniform $B_y$ Simulations (Standard Box Size and
 Resolution\tablenotemark{a} )}
\tablehead{
\colhead{Model} & 
\colhead{$\beta_0$} & 
\colhead{$Re_{M0}$} & \colhead{$X_0$} & 
\colhead{$\lambda_{\rm MRI}/H$\tablenotemark{b}} & 
\colhead{Orbits} &
\colhead{$\langle\negthinspace\langle w_M 
\rangle\negthinspace\rangle/P_0$} & 
\colhead{$\langle\negthinspace\langle w_R
\rangle\negthinspace\rangle/P_0$} & 
\colhead{$\langle\negthinspace\langle v_{{\rm A}z}^2 / \eta \Omega
\rangle\negthinspace\rangle$} 
}
\startdata
Y1 & 100 & 1000 & 0 & 0.63 & 100 &
0.0356 & 0.00852 & 293 \nl
Y2 & 100 & 100 & 0 & 0.63 & 100 &
0.0353 & 0.00833 & 28.6 \nl
Y3 & 100 & 100 & 0.2 & 0.63 & 100 &
0.0364 & 0.00870 & 32.1 \nl
Y4 & 100 & 100 & 0.4 & 0.63 & 100 &
0.0541 & 0.0125 & 53.2 \nl
Y5 & 100 & 30 & 0 & 0.63 & 100 &
0.0256 & 0.00616 & 6.14 \nl
Y6 & 100 & 30 & 0.2 & 0.63 & 100 &
0.0264 & 0.00626 & 6.53 \nl
Y7 & 100 & 30 & 0.4 & 0.63 & 100 &
0.0325 & 0.00739 & 8.84 \nl
Y8 & 100 & 10 & 0 & 0.63 & 100 &
$2.23 \times 10^{-5}$ & $5.95 \times 10^{-6}$ & 0.00246 \nl
Y9 & 100 & 10 & 0.2 & 0.63 & 100 &
$2.79 \times 10^{-5}$ & $8.88 \times 10^{-6}$ & 0.00331 \nl
Y10 & 100 & 10 & 0.4 & 0.63 & 100 &
$4.39 \times 10^{-5}$ & $1.48 \times 10^{-5}$ & 0.00636 \nl
Y11 & 100 & 10 & 10 & 0.63 & 200 &
$3.41 \times 10^{-7}$ & \nodata & $6.57 \times 10^{-4}$ \nl
Y12 & 100 & 3 & 0 & 0.63 & 100 &
\nodata & \nodata & \nodata \nl
Y13 & 100 & 3 & 0.2 & 0.63 & 100 &
\nodata & \nodata & \nodata \nl
Y14 & 100 & 3 & 0.4 & 0.63 & 100 &
\nodata & \nodata & \nodata \nl
Y15 & 100 & 3 & 10 & 0.63 & 100 &
\nodata & \nodata & \nodata \nl
Y16 & 100 & 3 & 100 & 0.63 & 100 &
\nodata & \nodata & \nodata \nl
Y17 & 400 & 1000 & 0 & 0.31 & 200 &
0.0279 & 0.00606 & 816 \nl
Y18 & 400 & 100 & 0 & 0.31 & 200 &
0.0286 & 0.00616 & 84.6 \nl
Y19 & 400 & 30 & 0 & 0.31 & 200 &
0.0261 & 0.00564 & 21.4 \nl
Y20 & 400 & 10 & 0 & 0.31 & 200 &
$5.34 \times 10^{-6}$ & $2.98 \times 10^{-6}$ & 
$5.45 \times 10^{-4}$ \nl
Y21 & 400 & 3 & 0 & 0.31 & 200 &
$5.59 \times 10^{-7}$ & $1.28 \times 10^{-7}$ & 
$7.83 \times 10^{-6}$ \nl
Y22 & 1600 & 1000 & 0 & 0.16 & 200 &
0.0223 & 0.00541 & $2.50 \times 10^3$ \nl
Y23 & 1600 & 100 & 0 & 0.16 & 200 &
0.0189 & 0.00467 & 200 \nl
Y24 & 1600 & 30 & 0 & 0.16 & 400 &
0.0234 & 0.00512 & 77.5 \nl
Y25 & 1600 & 10 & 0 & 0.16 & 200 &
$3.46 \times 10^{-6}$ & $ 4.51 \times 10^{-6}$ & 
$4.32 \times 10^{-4}$ \nl
Y26 & 1600 & 3 & 0 & 0.16 & 200 &
$2.20 \times 10^{-6}$ & $4.13 \times 10^{-6}$ & 
$8.51 \times 10^{-5}$ \nl
\enddata
\label{tbl:y}
\tablenotetext{a}{Box size is $H \times 4H \times H$ and grid
 resolutions is $32 \times 128 \times 32$.}
\tablenotetext{b}{$\lambda_{\rm MRI} \equiv 2 \pi v_{{\rm A}0} / \Omega$}
\end{deluxetable}

\begin{deluxetable}{cllllc}
\small
\tablewidth{0pc} 
\tablecaption{Time- and Volume-Averaged Values in Uniform $B_y$
 Simulations}
\tablehead{
\colhead{Quantity} & 
\colhead{Y2} & \colhead{Y4} & \colhead{Y8} & 
\colhead{Y10} & \colhead{Average\tablenotemark{a}}
}
\startdata
$\beta_0$ & 100 & 100 & 100 & 100 & \nodata \nl
$Re_{M0}$ & 100 & 100 & 10 & 10 & \nodata \nl
$X_{0}$ & 0 & 0.4 & 0 & 0.4 & \nodata \nl
$\langle\negthinspace\langle B_x^2 / 8 \pi 
\rangle\negthinspace\rangle / P_0$ &  
0.00890 & 0.0141 & $3.41 \times 10^{-6}$ & $5.63 \times 10^{-6}$ & 
$0.00743 \pm 0.00240 $\nl
$\langle\negthinspace\langle B_y^2 / 8 \pi 
\rangle\negthinspace\rangle / P_0$ &  
0.0702 & 0.109 & 0.00907 & 0.00943 & 
$0.0610 \pm 0.0177$ \nl
$\langle\negthinspace\langle B_z^2 / 8 \pi 
\rangle\negthinspace\rangle / P_0$ &  
0.00288 & 0.00538 & $2.44 \times 10^{-6}$ & $6.32 \times 10^{-6}$ &
$0.00247 \pm 0.00102$ \nl
$\langle\negthinspace\langle \rho v_x^2 / 2
\rangle\negthinspace\rangle / P_0$ &  
0.0139 & 0.0193 & $8.98 \times 10^{-6}$ &  $3.10 \times 10^{-5}$ &
$0.0117 \pm 0.0029$ \nl
$\langle\negthinspace\langle \rho \delta v_y^2 / 2
\rangle\negthinspace\rangle / P_0$ &  
0.0105 & 0.0157 & $2.42 \times 10^{-5}$ & $3.66 \times 10^{-5}$ &
$0.00872 \pm 0.00259$ \nl
$\langle\negthinspace\langle \rho v_z^2 / 2
\rangle\negthinspace\rangle / P_0$ &  
0.00583 & 0.00833 & $1.20 \times 10^{-5}$ & $3.84 \times 10^{-5}$ &
$0.00490 \pm 0.00134$ \nl
$\langle\negthinspace\langle P
\rangle\negthinspace\rangle / P_0$ &  
6.44 & 5.83 & 1.01 & 1.02 & 
\nodata \nl
$\langle\negthinspace\langle w_M
\rangle\negthinspace\rangle / 
\langle\negthinspace\langle w_R
\rangle\negthinspace\rangle $ &  
4.23 & 4.32 & 3.74 & 2.95 & 
$4.33 \pm 0.20$ \nl
$\langle\negthinspace\langle w_M
\rangle\negthinspace\rangle / 
\langle\negthinspace\langle B^2 / 8 \pi
\rangle\negthinspace\rangle $ &  
0.447 & 0.433 & 0.00245 & 0.00465 & 
$0.442 \pm 0.022$ \nl
$\langle\negthinspace\langle P
\rangle\negthinspace\rangle / 
\langle\negthinspace\langle B^2 / 8 \pi
\rangle\negthinspace\rangle $ &  
81.7 & 46.6 & 111 & 108 &
\nodata \nl
$\langle\negthinspace\langle \rho \delta v^2 / 2
\rangle\negthinspace\rangle / 
\langle\negthinspace\langle B^2 / 8 \pi
\rangle\negthinspace\rangle $ &  
0.383 & 0.347 & 0.00498 & 0.0112 & 
$0.372 \pm 0.018$ \nl
$\langle\negthinspace\langle B_x^2
\rangle\negthinspace\rangle / 
\langle\negthinspace\langle B_z^2
\rangle\negthinspace\rangle $ &  
3.09 & 2.61 & 1.40 & 0.891 &
$3.10 \pm 0.28$ \nl
$\langle\negthinspace\langle B_y^2
\rangle\negthinspace\rangle / 
\langle\negthinspace\langle B_z^2
\rangle\negthinspace\rangle $ &  
24.4 & 20.3 & $3.73 \times 10^3$ & $1.49 \times 10^3$ & 
$25.6 \pm 2.5$ \nl
$\langle\negthinspace\langle v_x^2
\rangle\negthinspace\rangle / 
\langle\negthinspace\langle v_z^2
\rangle\negthinspace\rangle $ &  
2.38 & 2.31 & 0.747 & 0.806 &
$2.42 \pm 0.13$ \nl
$\langle\negthinspace\langle \delta v_y^2
\rangle\negthinspace\rangle / 
\langle\negthinspace\langle v_z^2
\rangle\negthinspace\rangle $ &  
1.80 & 1.89 & 2.02 & 0.953 &
$1.78 \pm 0.10$ \nl
$\langle\negthinspace\langle \beta
\rangle\negthinspace\rangle$ &  
$1.39 \times 10^3$ & 688 & 114 & 115 &
\nodata \nl
$\langle\negthinspace\langle Re_M
\rangle\negthinspace\rangle$ &  
825 & $1.29 \times 10^3$ & 9.08 & 9.45 & \nodata \nl
$\langle\negthinspace\langle X_{\rm eff}
\rangle\negthinspace\rangle$ &  
0 & $-0.0112$ & 0 & $-0.00119$ & \nodata \nl
$\alpha$ &  
0.0436 & 0.0666 & $2.82 \times 10^{-5}$ & $5.88 \times 10^{-5}$ &
$0.0372 \pm 0.0106$ \nl
\enddata
\tablenotetext{a}{Models with $\langle \negthinspace \langle v_{{\rm A}z}^2 
/ \eta \Omega \rangle \negthinspace \rangle > 1$ are considered for 
the average (Y1 -- Y7, Y17 -- Y19, and Y22 -- Y24).}
\label{tbl:y2}
\end{deluxetable}

\clearpage

\begin{figure}
\begin{center}
\hspace*{-45pt}
\setlength{\unitlength}{0.1bp}%
{
}%
\begin{picture}(3600,2160)(0,0)%
{
}%
\put(3037,442){\makebox(0,0)[r]{\large{$-2$}}}%
\put(3037,599){\makebox(0,0)[r]{\large{$0$}}}%
\put(3037,756){\makebox(0,0)[r]{\large{$X_0=2$}}}%
\put(2200,50){\makebox(0,0){\large $t / t_{\rm rot}$}}%
\put(700,1180){%
\makebox(0,0)[b]{\shortstack{\large{$\log (\langle B^2 / 8 \pi \rangle / P_0)$}}}%
}%
\put(3450,200){\makebox(0,0)[c]{\large $50$}}%
\put(2950,200){\makebox(0,0)[c]{\large $40$}}%
\put(2450,200){\makebox(0,0)[c]{\large $30$}}%
\put(1950,200){\makebox(0,0)[c]{\large $20$}}%
\put(1450,200){\makebox(0,0)[c]{\large $10$}}%
\put(950,200){\makebox(0,0)[c]{\large $0$}}%
\put(900,1850){\makebox(0,0)[r]{\large $0$}}%
\put(900,1410){\makebox(0,0)[r]{\large $-1$}}%
\put(900,970){\makebox(0,0)[r]{\large $-2$}}%
\put(900,530){\makebox(0,0)[r]{\large $-3$}}%
\end{picture}%
\\ 
\caption
{Time evolution of the volume-averaged magnetic energy $\langle B^2 / 8
\pi \rangle / P_0$ for models Z2 ($X_0 = 2$), Z3 ($X_0 = 0$), and Z4
($X_0 = -2$).
The plasma beta and the magnetic Reynolds number of these models are
$\beta_0 = 3200$ and $Re_{M0} = 100$.
\label{fig:em-t-z}}
\end{center}
\end{figure}

\begin{figure}
\begin{center}
\hspace*{-45pt}
\setlength{\unitlength}{0.1bp}%
{
}%
\begin{picture}(3600,2160)(0,0)%
{
}%
\put(3037,1605){\makebox(0,0)[r]{\large{$-2$}}}%
\put(3037,1762){\makebox(0,0)[r]{\large{$0$}}}%
\put(3037,1919){\makebox(0,0)[r]{\large{$X_0=2$}}}%
\put(2225,50){\makebox(0,0){\large $t / t_{\rm rot}$}}%
\put(700,1180){%
\makebox(0,0)[b]{\shortstack{\large{$\langle - B_x B_y / 4 \pi \rangle / P_0$}}}%
}%
\put(3450,200){\makebox(0,0)[c]{\large $50$}}%
\put(2960,200){\makebox(0,0)[c]{\large $40$}}%
\put(2470,200){\makebox(0,0)[c]{\large $30$}}%
\put(1980,200){\makebox(0,0)[c]{\large $20$}}%
\put(1490,200){\makebox(0,0)[c]{\large $10$}}%
\put(1000,200){\makebox(0,0)[c]{\large $0$}}%
\put(950,2060){\makebox(0,0)[r]{\large $0.6$}}%
\put(950,1767){\makebox(0,0)[r]{\large $0.5$}}%
\put(950,1473){\makebox(0,0)[r]{\large $0.4$}}%
\put(950,1180){\makebox(0,0)[r]{\large $0.3$}}%
\put(950,887){\makebox(0,0)[r]{\large $0.2$}}%
\put(950,593){\makebox(0,0)[r]{\large $0.1$}}%
\put(950,300){\makebox(0,0)[r]{\large $0$}}%
\end{picture}%
\\ 
\caption
{Time evolution of the volume-averaged Maxwell stress $\langle - B_x B_y
/ 4 \pi \rangle / P_0$ for models Z2 ($X_0 = 2$), Z3 ($X_0 = 0$), and Z4
($X_0 = -2$).
The plasma beta and the magnetic Reynolds number of these models are
$\beta_0 = 3200$ and $Re_{M0} = 100$.
\label{fig:wm-t-z}}
\end{center}
\end{figure}

\clearpage

\begin{figure}
\caption
{Slices in the $x$-$z$ plane at $y = - 2 H$ and in the $y$-$z$ plane at
$x = - 0.5 H$ of the magnetic energy, $\log (B^2 / 8 \pi P_0)$
($colors$), and perturbed velocity field, $\delta v$ ($arrows$), in
model Z2 ($\beta_0 = 3200$, $Re_{M0} = 100$, and $X_0 = 2$) at 18, 21,
and 25 orbits. 
\label{fig:z2}}
\end{figure}

\begin{figure}
\caption
{Slices in the $x$-$z$ plane at $y = - 2 H$ and in the $y$-$z$ plane at
$x = - 0.5 H$ of the magnetic energy, $\log (B^2 / 8 \pi P_0)$
($colors$), and perturbed velocity field, $\delta v$ ($arrows$), in
model Z4 ($\beta_0 = 3200$, $Re_{M0} = 100$, and $X_0 = - 2$) at 10, 17,
and 25 orbits.
\label{fig:z4}}
\end{figure}

\clearpage

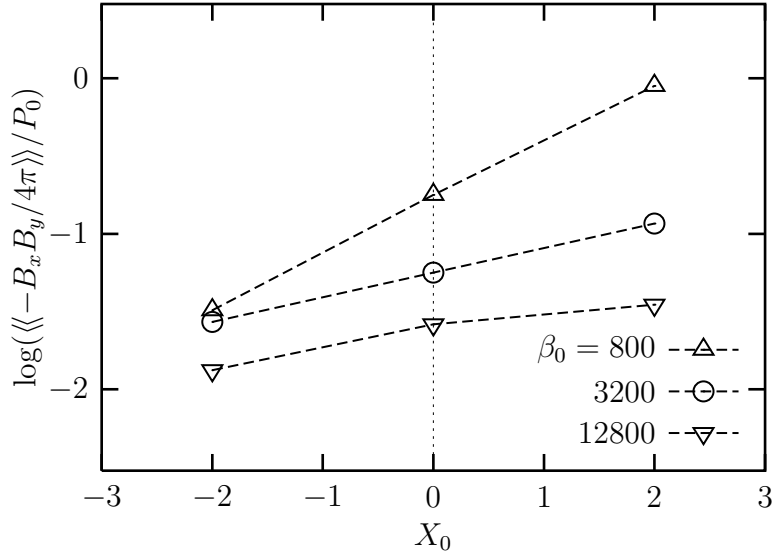
\begin{figure}
\begin{center}
\hspace*{-45pt}
\setlength{\unitlength}{0.1bp}%
{
}%
\begin{picture}(3600,2160)(0,0)%
{
}%
\put(3037,442){\makebox(0,0)[r]{\large{$12800$}}}%
\put(3037,599){\makebox(0,0)[r]{\large{$3200$}}}%
\put(3037,756){\makebox(0,0)[r]{\large{$\beta_0 = 800$}}}%
\put(2200,50){\makebox(0,0){\large $X_0$}}%
\put(700,1180){%
\makebox(0,0)[b]{\shortstack{\large{$\log (\langle \negthinspace \langle - B_x B_y / 4 \pi \rangle \negthinspace \rangle / P_{0})$}}}%
}%
\put(3450,200){\makebox(0,0)[c]{\large $3$}}%
\put(3033,200){\makebox(0,0)[c]{\large $2$}}%
\put(2617,200){\makebox(0,0)[c]{\large $1$}}%
\put(2200,200){\makebox(0,0)[c]{\large $0$}}%
\put(1783,200){\makebox(0,0)[c]{\large $-1$}}%
\put(1367,200){\makebox(0,0)[c]{\large $-2$}}%
\put(950,200){\makebox(0,0)[c]{\large $-3$}}%
\put(900,1780){\makebox(0,0)[r]{\large $0$}}%
\put(900,1193){\makebox(0,0)[r]{\large $-1$}}%
\put(900,607){\makebox(0,0)[r]{\large $-2$}}%
\end{picture}%
\\ 
\caption
{Saturation level of the Maxwell stress as a function of the Hall
parameter $X_0$ for the models with $\beta_0 = 800$, 3200, and 12800.
The magnetic Reynolds number is $Re_{M0} = 1$ for all the models.
\label{fig:wm-x-rm1-z}}
\end{center}
\end{figure}

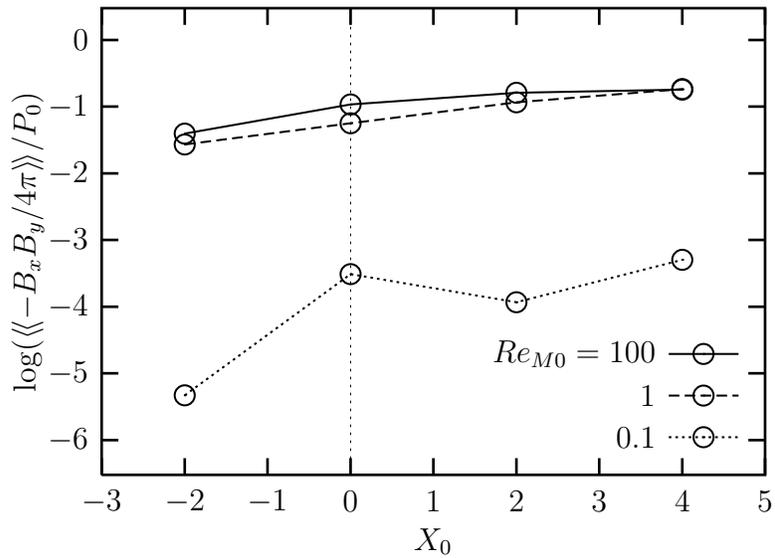
\begin{figure}
\begin{center}
\hspace*{-45pt}
\setlength{\unitlength}{0.1bp}%
{
}%
\begin{picture}(3600,2160)(0,0)%
{
}%
\put(3037,442){\makebox(0,0)[r]{\large{$0.1$}}}%
\put(3037,599){\makebox(0,0)[r]{\large{$1$}}}%
\put(3037,756){\makebox(0,0)[r]{\large{$Re_{M0} = 100$}}}%
\put(2200,50){\makebox(0,0){\large $X_0$}}%
\put(700,1180){%
\makebox(0,0)[b]{\shortstack{\large{$\log (\langle \negthinspace \langle - B_x B_y / 4 \pi \rangle \negthinspace \rangle / P_{0})$}}}%
}%
\put(3450,200){\makebox(0,0)[c]{\large $5$}}%
\put(3138,200){\makebox(0,0)[c]{\large $4$}}%
\put(2825,200){\makebox(0,0)[c]{\large $3$}}%
\put(2513,200){\makebox(0,0)[c]{\large $2$}}%
\put(2200,200){\makebox(0,0)[c]{\large $1$}}%
\put(1888,200){\makebox(0,0)[c]{\large $0$}}%
\put(1575,200){\makebox(0,0)[c]{\large $-1$}}%
\put(1263,200){\makebox(0,0)[c]{\large $-2$}}%
\put(950,200){\makebox(0,0)[c]{\large $-3$}}%
\put(900,1940){\makebox(0,0)[r]{\large $0$}}%
\put(900,1689){\makebox(0,0)[r]{\large $-1$}}%
\put(900,1437){\makebox(0,0)[r]{\large $-2$}}%
\put(900,1186){\makebox(0,0)[r]{\large $-3$}}%
\put(900,934){\makebox(0,0)[r]{\large $-4$}}%
\put(900,683){\makebox(0,0)[r]{\large $-5$}}%
\put(900,431){\makebox(0,0)[r]{\large $-6$}}%
\end{picture}%
\\ 
\caption
{Saturation level of the Maxwell stress as a function of the Hall
parameter $X_0$ for the models with $Re_{M0} = 100$, 1, and 0.1.
The plasma beta is $\beta_0 = 3200$ for all the models.
\label{fig:wm-x-b32-z}}
\end{center}
\end{figure}

\clearpage

\begin{figure}
\begin{center}
\begin{minipage}{.4\textwidth}
\hspace*{-22.5pt}
\setlength{\unitlength}{0.1bp}%
{
}%
\begin{picture}(2394,2160)(0,0)%
{
}%
\put(1261,429){\makebox(0,0)[r]{\large{$k_z$}}}%
\put(1261,547){\makebox(0,0)[r]{\large{$k_y$}}}%
\put(1261,665){\makebox(0,0)[r]{\large{$k_x$}}}%
\put(1828,1759){\makebox(0,0)[l]{\large $k^{-4}$}}%
\put(1105,1934){\makebox(0,0)[l]{\large (a) $X_0 = 2$}}%
\put(1622,50){\makebox(0,0){\large $k H / 2 \pi$}}%
\put(700,1180){%
\makebox(0,0)[b]{\shortstack{\large{$\log (|\negthinspace \mbox{\boldmath{$B$}}(\mbox{\boldmath{$k$}})|^2 / 8 \pi P_0)$}}}%
}%
\put(1983,200){\makebox(0,0)[c]{\large $10$}}%
\put(1467,200){\makebox(0,0)[c]{\large $1$}}%
\put(950,2060){\makebox(0,0)[r]{\large $0$}}%
\put(950,1809){\makebox(0,0)[r]{\large $-2$}}%
\put(950,1557){\makebox(0,0)[r]{\large $-4$}}%
\put(950,1306){\makebox(0,0)[r]{\large $-6$}}%
\put(950,1054){\makebox(0,0)[r]{\large $-8$}}%
\put(950,803){\makebox(0,0)[r]{\large $-10$}}%
\put(950,551){\makebox(0,0)[r]{\large $-12$}}%
\put(950,300){\makebox(0,0)[r]{\large $-14$}}%
\end{picture}%
\end{minipage}
\begin{minipage}{.4\textwidth}
\hspace*{-85pt}
\setlength{\unitlength}{0.1bp}%
{
}%
\begin{picture}(2394,2160)(0,0)%
{
}%
\put(1224,429){\makebox(0,0)[r]{\large{$k_z$}}}%
\put(1224,547){\makebox(0,0)[r]{\large{$k_y$}}}%
\put(1224,665){\makebox(0,0)[r]{\large{$k_x$}}}%
\put(1811,1595){\makebox(0,0)[l]{\large $k^{-4}$}}%
\put(1060,1934){\makebox(0,0)[l]{\large (b) $X_0 = -2$}}%
\put(1597,50){\makebox(0,0){\large $k H / 2 \pi$}}%
\put(700,1180){%
\makebox(0,0)[b]{\shortstack{ }}%
}%
\put(1973,200){\makebox(0,0)[c]{\large $10$}}%
\put(1435,200){\makebox(0,0)[c]{\large $1$}}%
\put(900,2060){\makebox(0,0)[r]{ }}%
\put(900,2060){\makebox(0,0)[r]{ }}%
\put(900,1809){\makebox(0,0)[r]{ }}%
\put(900,1557){\makebox(0,0)[r]{ }}%
\put(900,1306){\makebox(0,0)[r]{ }}%
\put(900,1054){\makebox(0,0)[r]{ }}%
\put(900,803){\makebox(0,0)[r]{ }}%
\put(900,551){\makebox(0,0)[r]{ }}%
\put(900,300){\makebox(0,0)[r]{\phantom{aaaaaa}}}%
\end{picture}%
\end{minipage}
\caption
{Power spectra of magnetic field in models (a) Z2 and (b) Z4 along the
$k_x$, $k_y$, and $k_z$ axes.
The wavenumbers are normalized in terms of $H / 2 \pi$ and energies in
terms of initial pressure $P_0$.
\label{fig:spd}}
\end{center}
\end{figure}
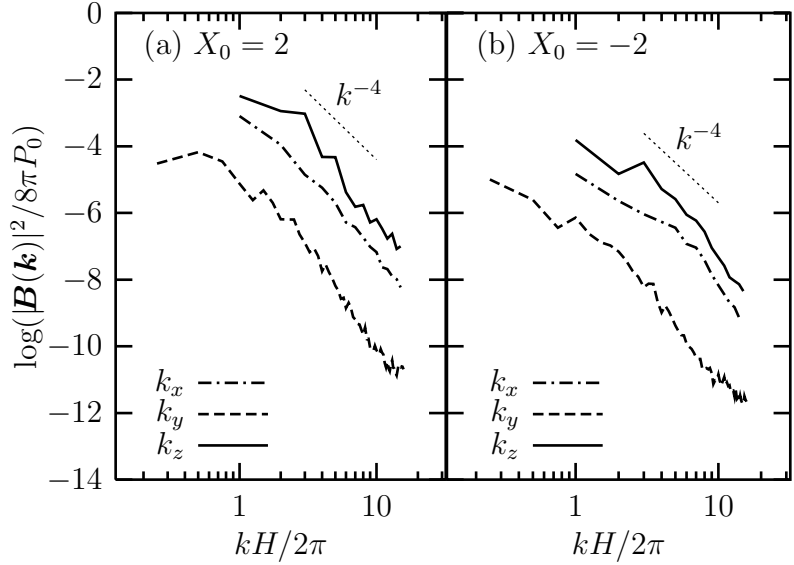

\begin{figure}
\begin{center}
\begin{minipage}{.4\textwidth}
\hspace*{-22.5pt}
\setlength{\unitlength}{0.1bp}%
{
}%
\begin{picture}(2394,2160)(0,0)%
{
}%
\put(1261,426){\makebox(0,0)[r]{\large{$k_z$}}}%
\put(1261,544){\makebox(0,0)[r]{\large{$k_y$}}}%
\put(1261,662){\makebox(0,0)[r]{\large{$k_x$}}}%
\put(1598,1698){\makebox(0,0)[l]{\large $k_{\rm crit}$}}%
\put(1402,1698){\makebox(0,0)[r]{\large $k_{\rm diff}$}}%
\put(1861,1405){\makebox(0,0)[l]{\large $k^{-8}$}}%
\put(1105,1933){\makebox(0,0)[l]{\large (a) $| \mbox{\boldmath{$B$}}(\mbox{\boldmath{$k$}})|^2 / 8 \pi$}}%
\put(1622,50){\makebox(0,0){\large $k H / 2 \pi$}}%
\put(700,1180){%
\makebox(0,0)[b]{\shortstack{\large{$\log (| E (\mbox{\boldmath{$k$}})| / P_0)$}}}%
}%
\put(1983,200){\makebox(0,0)[c]{\large $10$}}%
\put(1467,200){\makebox(0,0)[c]{\large $1$}}%
\put(950,2060){\makebox(0,0)[r]{\large $0$}}%
\put(950,1864){\makebox(0,0)[r]{\large $-2$}}%
\put(950,1669){\makebox(0,0)[r]{\large $-4$}}%
\put(950,1473){\makebox(0,0)[r]{\large $-6$}}%
\put(950,1278){\makebox(0,0)[r]{\large $-8$}}%
\put(950,1082){\makebox(0,0)[r]{\large $-10$}}%
\put(950,887){\makebox(0,0)[r]{\large $-12$}}%
\put(950,691){\makebox(0,0)[r]{\large $-14$}}%
\put(950,496){\makebox(0,0)[r]{\large $-16$}}%
\put(950,300){\makebox(0,0)[r]{\large $-18$}}%
\end{picture}%
\end{minipage}
\begin{minipage}{.4\textwidth}
\hspace*{-85pt}
\setlength{\unitlength}{0.1bp}%
{
}%
\begin{picture}(2394,2160)(0,0)%
{
}%
\put(1224,426){\makebox(0,0)[r]{\large{$k_z$}}}%
\put(1224,544){\makebox(0,0)[r]{\large{$k_y$}}}%
\put(1224,662){\makebox(0,0)[r]{\large{$k_x$}}}%
\put(1530,1112){\makebox(0,0)[r]{\large $k^{-8}$}}%
\put(1060,1933){\makebox(0,0)[l]{\large (b) $ \rho_0 | \mbox{\boldmath{$v$}}(\mbox{\boldmath{$k$}})|^2 / 2$}}%
\put(1597,50){\makebox(0,0){\large $k H / 2 \pi$}}%
\put(100,1530){%
\makebox(0,0)[b]{\shortstack{ }}%
}%
\put(1973,200){\makebox(0,0)[c]{\large $10$}}%
\put(1435,200){\makebox(0,0)[c]{\large $1$}}%
\put(900,2060){\makebox(0,0)[r]{ }}%
\put(900,1864){\makebox(0,0)[r]{ }}%
\put(900,1669){\makebox(0,0)[r]{ }}%
\put(900,1473){\makebox(0,0)[r]{ }}%
\put(900,1278){\makebox(0,0)[r]{ }}%
\put(900,1082){\makebox(0,0)[r]{ }}%
\put(900,887){\makebox(0,0)[r]{ }}%
\put(900,691){\makebox(0,0)[r]{ }}%
\put(900,496){\makebox(0,0)[r]{ }}%
\put(900,300){\makebox(0,0)[r]{\phantom{aaaa}}}%
\end{picture}%
\end{minipage}
\caption
{Power spectra of (a) magnetic field and (b) velocity in model Z13 along
the $k_x$, $k_y$, and $k_z$ axes.
The wavenumbers are normalized in terms of $H / 2 \pi$ and energies in
terms of initial pressure $P_0$.
The arrows lie at the diffusion scale $k_{\rm diff} \equiv v_{{\rm A}0}
/ \eta$ and at the critical wavenumber for the MRI $k_{\rm crit}$.
\label{fig:spd2}}
\end{center}
\end{figure}
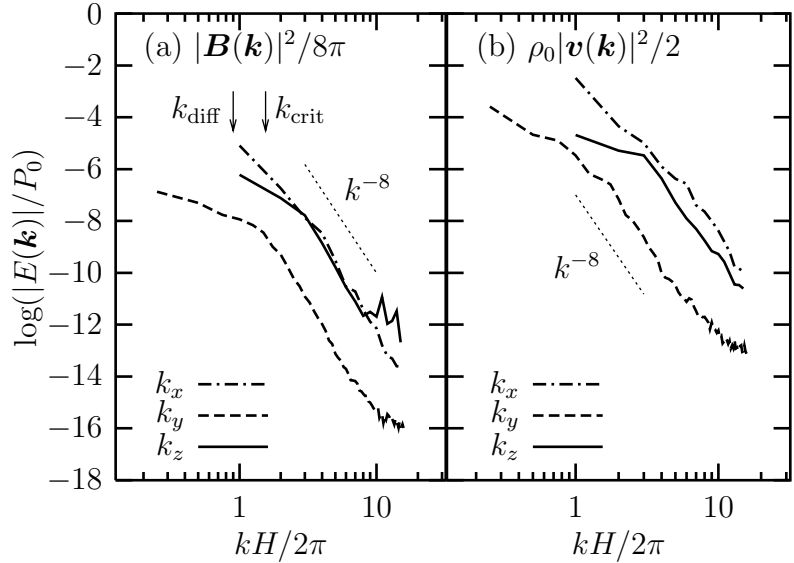

\clearpage

\begin{figure}
\begin{center}
\hspace*{-45pt}
\setlength{\unitlength}{0.1bp}%
{
}%
\begin{picture}(3600,2160)(0,0)%
{
}%
\put(3037,442){\makebox(0,0)[r]{\large{$3200$, \hspace{20.7pt} $-2$}}}%
\put(3037,599){\makebox(0,0)[r]{\large{$3200$, \hspace{31.8pt} $4$}}}%
\put(3037,756){\makebox(0,0)[r]{\large{$12800$, \hspace{31.8pt} $0$}}}%
\put(3037,913){\makebox(0,0)[r]{\large{$\beta_0 = 3200$, $X_0 = 0$}}}%
\put(2200,50){\makebox(0,0){\large $Re_{M0}$}}%
\put(700,1180){%
\makebox(0,0)[b]{\shortstack{\large{$\log (\langle \negthinspace \langle - B_x B_y / 4 \pi \rangle \negthinspace \rangle / P_{0})$}}}%
}%
\put(3152,200){\makebox(0,0)[c]{\large $100$}}%
\put(2527,200){\makebox(0,0)[c]{\large $10$}}%
\put(1902,200){\makebox(0,0)[c]{\large $1$}}%
\put(1277,200){\makebox(0,0)[c]{\large $0.1$}}%
\put(900,2060){\makebox(0,0)[r]{\large $0$}}%
\put(900,1767){\makebox(0,0)[r]{\large $-1$}}%
\put(900,1473){\makebox(0,0)[r]{\large $-2$}}%
\put(900,1180){\makebox(0,0)[r]{\large $-3$}}%
\put(900,887){\makebox(0,0)[r]{\large $-4$}}%
\put(900,593){\makebox(0,0)[r]{\large $-5$}}%
\put(900,300){\makebox(0,0)[r]{\large $-6$}}%
\end{picture}%
\\ 
\caption
{Saturation level of the Maxwell stress as a function of the magnetic
Reynolds number $Re_{M0}$.
Open circles and triangles denotes the models without Hall term ($X_0 =
0$) for $\beta_0 = 3200$ and $\beta_0 = 12800$, respectively.
The models including the Hall term are shown by filled circles ($X_0 =
4$) and triangles ($X_0 = - 2$).
\label{fig:wm-rm-z}}
\end{center}
\end{figure}
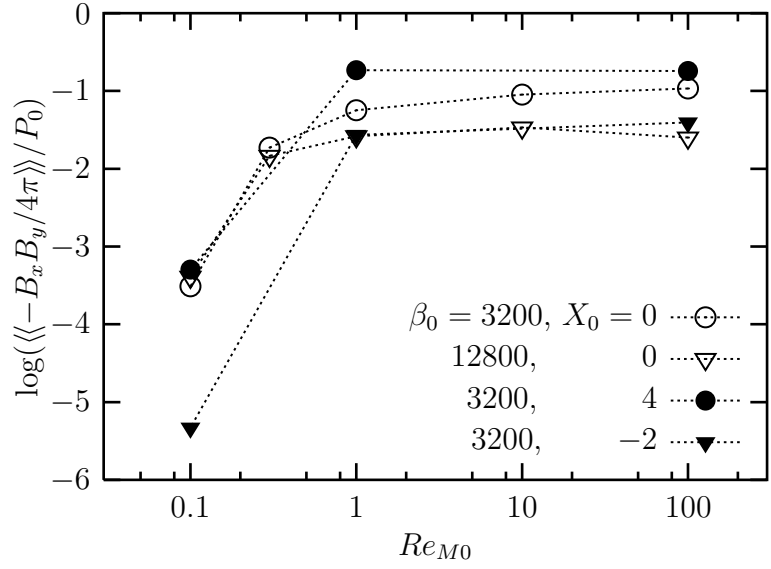

\clearpage

\begin{figure}
\begin{center}
\hspace*{-45pt}
\setlength{\unitlength}{0.1bp}%
{
}%
\begin{picture}(3600,2160)(0,0)%
{
}%
\put(1576,1640){\makebox(0,0)[r]{\large{$4$}}}%
\put(1576,1797){\makebox(0,0)[r]{\large{$2$}}}%
\put(1576,1954){\makebox(0,0)[r]{\large{$X_0=0$}}}%
\put(2250,50){\makebox(0,0){\large $t / t_{\rm rot}$}}%
\put(700,1180){%
\makebox(0,0)[b]{\shortstack{\large{$\langle - B_x B_y / 4 \pi \rangle / P_0$}}}%
}%
\put(3450,200){\makebox(0,0)[c]{\large $50$}}%
\put(2970,200){\makebox(0,0)[c]{\large $40$}}%
\put(2490,200){\makebox(0,0)[c]{\large $30$}}%
\put(2010,200){\makebox(0,0)[c]{\large $20$}}%
\put(1530,200){\makebox(0,0)[c]{\large $10$}}%
\put(1050,200){\makebox(0,0)[c]{\large $0$}}%
\put(1000,2060){\makebox(0,0)[r]{\large $0.14$}}%
\put(1000,1809){\makebox(0,0)[r]{\large $0.12$}}%
\put(1000,1557){\makebox(0,0)[r]{\large $0.1$}}%
\put(1000,1306){\makebox(0,0)[r]{\large $0.08$}}%
\put(1000,1054){\makebox(0,0)[r]{\large $0.06$}}%
\put(1000,803){\makebox(0,0)[r]{\large $0.04$}}%
\put(1000,551){\makebox(0,0)[r]{\large $0.02$}}%
\put(1000,300){\makebox(0,0)[r]{\large $0$}}%
\end{picture}%
\\ 
\caption
{Time evolution of the volume-averaged Maxwell stress $\langle - B_x B_y
/ 4 \pi \rangle / P_0$ for models S1 ($X_0 = 0$), S2 ($X_0 = 2$), and S3
($X_0 = 4$).
The plasma beta and the magnetic Reynolds number of these models are
$\beta_0 = 3200$ and $Re_{M0} = 100$.
\label{fig:wm-t-s}}
\end{center}
\end{figure}

\begin{figure}
\caption
{Slices in the $x$-$z$ plane at $y = 0$ of the azimuthal component of
the magnetic energy, $\log (B_y^2 / 8 \pi P_0)$ ($colors$), and
magnetic field vector, $B$ ($arrows$), in model S3 ($\beta_0 = 3200$,
$Re_{M0} = 100$, and $X_0 = 4$) at 35, 40, 45, and 50 orbits.
\label{fig:s3}}
\end{figure}

\clearpage

\begin{figure}
\begin{center}
\hspace*{-45pt}
\setlength{\unitlength}{0.1bp}%
{
}%
\begin{picture}(3600,2160)(0,0)%
{
}%
\put(3025,992){\makebox(0,0)[r]{\large{$4$}}}%
\put(3025,1149){\makebox(0,0)[r]{\large{$2$}}}%
\put(3025,1306){\makebox(0,0)[r]{\large{$X_0=0$}}}%
\put(2200,50){\makebox(0,0){\large $t / t_{\rm rot}$}}%
\put(700,1180){%
\makebox(0,0)[b]{\shortstack{\large{$\log (\langle B^2 / 8 \pi \rangle / P_0)$}}}%
}%
\put(3450,200){\makebox(0,0)[c]{\large $50$}}%
\put(2950,200){\makebox(0,0)[c]{\large $40$}}%
\put(2450,200){\makebox(0,0)[c]{\large $30$}}%
\put(1950,200){\makebox(0,0)[c]{\large $20$}}%
\put(1450,200){\makebox(0,0)[c]{\large $10$}}%
\put(950,200){\makebox(0,0)[c]{\large $0$}}%
\put(900,1850){\makebox(0,0)[r]{\large $-1$}}%
\put(900,1410){\makebox(0,0)[r]{\large $-2$}}%
\put(900,970){\makebox(0,0)[r]{\large $-3$}}%
\put(900,530){\makebox(0,0)[r]{\large $-4$}}%
\end{picture}%
\\ 
\caption
{Time evolution of the volume-averaged magnetic energy $\langle B^2 / 8
\pi \rangle / P_0$ for models S8 ($X_0 = 0$), S9 ($X_0 = 2$), and S10
($X_0 = 4$).
The plasma beta and the magnetic Reynolds number of these models are
$\beta_0 = 3200$ and $Re_{M0} = 1$.
\label{fig:em-t-s}}
\end{center}
\end{figure}

\begin{figure}
\begin{center}
\hspace*{-45pt}
\setlength{\unitlength}{0.1bp}%
{
}%
\begin{picture}(3600,2160)(0,0)%
{
}%
\put(3037,442){\makebox(0,0)[r]{\large{$12800$}}}%
\put(3037,599){\makebox(0,0)[r]{\large{$3200$}}}%
\put(3037,756){\makebox(0,0)[r]{\large{$\beta_0=800$}}}%
\put(2200,50){\makebox(0,0){\large $Re_{M0}$}}%
\put(700,1180){%
\makebox(0,0)[b]{\shortstack{\large{$\log (\langle \negthinspace \langle - B_x B_y / 4 \pi \rangle \negthinspace \rangle / P_{0})$}}}%
}%
\put(3161,200){\makebox(0,0)[c]{\large $100$}}%
\put(2200,200){\makebox(0,0)[c]{\large $10$}}%
\put(1239,200){\makebox(0,0)[c]{\large $1$}}%
\put(900,1940){\makebox(0,0)[r]{\large $-1$}}%
\put(900,1689){\makebox(0,0)[r]{\large $-2$}}%
\put(900,1437){\makebox(0,0)[r]{\large $-3$}}%
\put(900,1186){\makebox(0,0)[r]{\large $-4$}}%
\put(900,934){\makebox(0,0)[r]{\large $-5$}}%
\put(900,683){\makebox(0,0)[r]{\large $-6$}}%
\put(900,431){\makebox(0,0)[r]{\large $-7$}}%
\end{picture}%
\\ 
\caption
{Saturation level of the Maxwell stress as a function of the magnetic
Reynolds number $Re_{M0}$ for zero-net flux $B_z$ models.
All the models are without the Hall effect ($X_0 = 0$), and the initial
field strength is $\beta_0 = 800$, 3200, and 12800.
\label{fig:wm-rm-s}}
\end{center}
\end{figure}
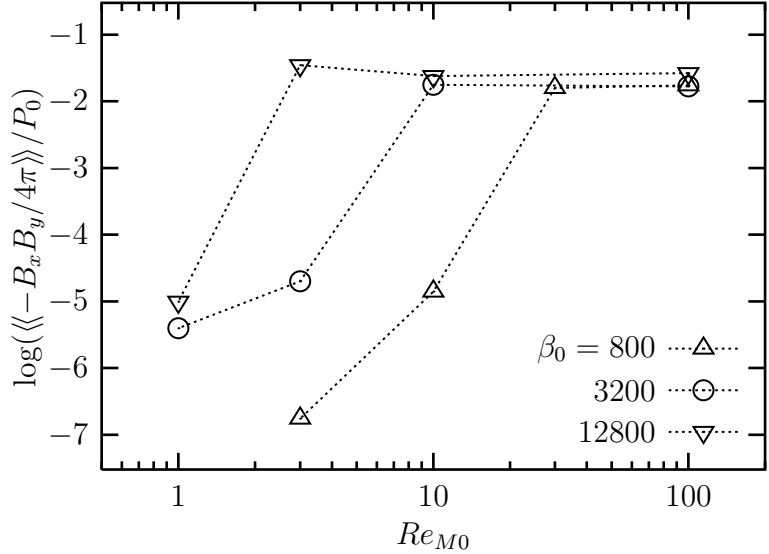

\clearpage

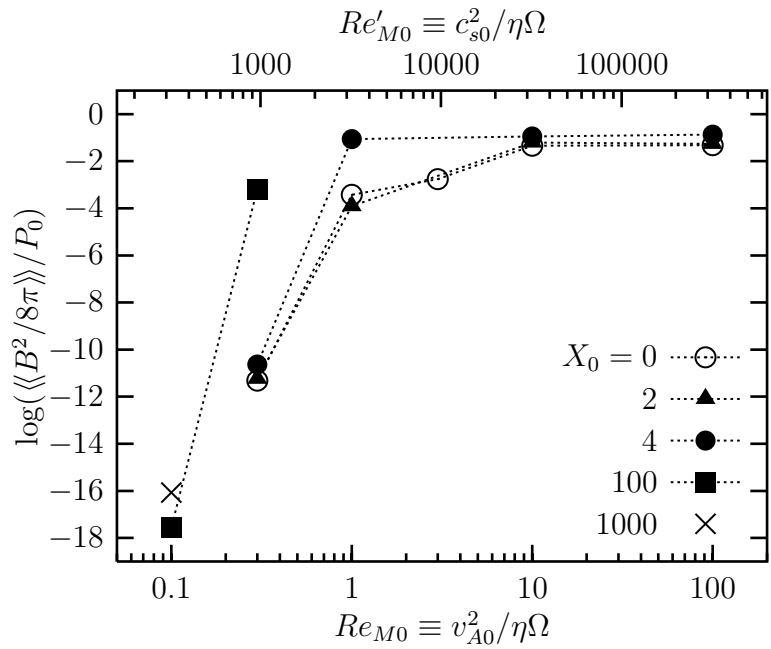
\begin{figure}
\begin{center}
\hspace*{-45pt}
\setlength{\unitlength}{0.1bp}%
{
}%
\begin{picture}(3600,2376)(0,0)%
{
}%
\put(3037,442){\makebox(0,0)[r]{\large{$1000$}}}%
\put(3037,599){\makebox(0,0)[r]{\large{$100$}}}%
\put(3037,756){\makebox(0,0)[r]{\large{$4$}}}%
\put(3037,913){\makebox(0,0)[r]{\large{$2$}}}%
\put(3037,1070){\makebox(0,0)[r]{\large{$X_{0} = 0$}}}%
\put(2225,2325){\makebox(0,0){\large $Re_{M0}' \equiv c_{s0}^2/\eta \Omega$}}%
\put(2225,50){\makebox(0,0){\large $Re_{M0} \equiv v_{A0}^2/\eta \Omega$}}%
\put(700,1188){%
\makebox(0,0)[b]{\shortstack{\large{$\log (\langle \negthinspace \langle B^2 / 8 \pi \rangle \negthinspace \rangle / P_{0})$}}}%
}%
\put(2902,2176){\makebox(0,0)[c]{\large $100000$}}%
\put(2221,2176){\makebox(0,0)[c]{\large $10000$}}%
\put(1541,2176){\makebox(0,0)[c]{\large $1000$}}%
\put(3245,200){\makebox(0,0)[c]{\large $100$}}%
\put(2565,200){\makebox(0,0)[c]{\large $10$}}%
\put(1885,200){\makebox(0,0)[c]{\large $1$}}%
\put(1205,200){\makebox(0,0)[c]{\large $0.1$}}%
\put(950,1987){\makebox(0,0)[r]{\large $0$}}%
\put(950,1810){\makebox(0,0)[r]{\large $-2$}}%
\put(950,1632){\makebox(0,0)[r]{\large $-4$}}%
\put(950,1454){\makebox(0,0)[r]{\large $-6$}}%
\put(950,1277){\makebox(0,0)[r]{\large $-8$}}%
\put(950,1099){\makebox(0,0)[r]{\large $-10$}}%
\put(950,922){\makebox(0,0)[r]{\large $-12$}}%
\put(950,744){\makebox(0,0)[r]{\large $-14$}}%
\put(950,566){\makebox(0,0)[r]{\large $-16$}}%
\put(950,389){\makebox(0,0)[r]{\large $-18$}}%
\end{picture}%
\\ 
\caption
{Saturation level of the magnetic energy as a function of the magnetic
Reynolds number $Re_{M0}$ for zero-net flux $B_z$ models ($\beta_0 =
3200$).
Open circles denote the models with only the ohmic dissipation ($X_0 =
0$), and the other marks are including also the Hall effect ($X_0 = 2$,
4, 100, and 1000).
\label{fig:em-rm-s}}
\end{center}
\end{figure}

\begin{figure}
\begin{center}
\hspace*{-45pt}
\setlength{\unitlength}{0.1bp}%
{
}%
\begin{picture}(3600,2160)(0,0)%
{
}%
\put(1576,1640){\makebox(0,0)[r]{\large{$0.4$}}}%
\put(1576,1797){\makebox(0,0)[r]{\large{$0.2$}}}%
\put(1576,1954){\makebox(0,0)[r]{\large{$X_0=0$}}}%
\put(2250,50){\makebox(0,0){\large $t / t_{\rm rot}$}}%
\put(700,1180){%
\makebox(0,0)[b]{\shortstack{\large{$\langle - B_x B_y / 4 \pi \rangle / P_0$}}}%
}%
\put(3450,200){\makebox(0,0)[c]{\large $100$}}%
\put(2970,200){\makebox(0,0)[c]{\large $80$}}%
\put(2490,200){\makebox(0,0)[c]{\large $60$}}%
\put(2010,200){\makebox(0,0)[c]{\large $40$}}%
\put(1530,200){\makebox(0,0)[c]{\large $20$}}%
\put(1050,200){\makebox(0,0)[c]{\large $0$}}%
\put(1000,2060){\makebox(0,0)[r]{\large $0.14$}}%
\put(1000,1809){\makebox(0,0)[r]{\large $0.12$}}%
\put(1000,1557){\makebox(0,0)[r]{\large $0.1$}}%
\put(1000,1306){\makebox(0,0)[r]{\large $0.08$}}%
\put(1000,1054){\makebox(0,0)[r]{\large $0.06$}}%
\put(1000,803){\makebox(0,0)[r]{\large $0.04$}}%
\put(1000,551){\makebox(0,0)[r]{\large $0.02$}}%
\put(1000,300){\makebox(0,0)[r]{\large $0$}}%
\end{picture}%
\\ 
\caption
{Time evolution of the volume-averaged Maxwell stress $\langle - B_x B_y
/ 4 \pi \rangle / P_0$ for models Y2 ($X_0 = 0$), Y3 ($X_0 = 0.2$), and Y4
($X_0 = 0.4$).
The plasma beta and the magnetic Reynolds number of these models are
$\beta_0 = 100$ and $Re_{M0} = 100$.
\label{fig:wm-t-y}}
\end{center}
\end{figure}

\begin{figure}
\caption
{Slices in the $x$-$z$ plane at $y = 0$ of the azimuthal component of
the magnetic energy, $\log (B_y^2 / 8 \pi P_0)$ ($colors$), and
perturbed velocity field $\delta v$ ($arrows$) in model Y4 ($\beta_0 =
100$, $Re_{M0} = 100$, and $X_0 = 0.4$) at 5, 10, 15, and 20 orbits.
\label{fig:y4}}
\end{figure}

\begin{figure}
\begin{center}
\hspace*{-45pt}
\setlength{\unitlength}{0.1bp}%
{
}%
\begin{picture}(3600,2160)(0,0)%
{
}%
\put(3033,606){\makebox(0,0)[r]{\large{$3$}}}%
\put(3033,763){\makebox(0,0)[r]{\large{$10$}}}%
\put(3033,920){\makebox(0,0)[r]{\large{$30$}}}%
\put(3033,1077){\makebox(0,0)[r]{\large{$Re_{M0}=100$}}}%
\put(2225,50){\makebox(0,0){\large $t / t_{\rm rot}$}}%
\put(700,1180){%
\makebox(0,0)[b]{\shortstack{\large{$\log (\langle (B_x^2+B_z^2) / 8 \pi \rangle / P_0)$}}}%
}%
\put(3450,200){\makebox(0,0)[c]{\large $100$}}%
\put(2960,200){\makebox(0,0)[c]{\large $80$}}%
\put(2470,200){\makebox(0,0)[c]{\large $60$}}%
\put(1980,200){\makebox(0,0)[c]{\large $40$}}%
\put(1490,200){\makebox(0,0)[c]{\large $20$}}%
\put(1000,200){\makebox(0,0)[c]{\large $0$}}%
\put(950,1950){\makebox(0,0)[r]{\large $0$}}%
\put(950,1730){\makebox(0,0)[r]{\large $-2$}}%
\put(950,1510){\makebox(0,0)[r]{\large $-4$}}%
\put(950,1290){\makebox(0,0)[r]{\large $-6$}}%
\put(950,1070){\makebox(0,0)[r]{\large $-8$}}%
\put(950,850){\makebox(0,0)[r]{\large $-10$}}%
\put(950,630){\makebox(0,0)[r]{\large $-12$}}%
\put(950,410){\makebox(0,0)[r]{\large $-14$}}%
\end{picture}%
\\ 
\caption
{Time evolution of the volume-averaged poloidal component of the
magnetic energy  $\langle (B_x^2  + B_z^2) / 8 \pi \rangle / P_0$ for
models Y4 ($Re_{M0} = 100$), Y7 ($Re_{M0} = 30$), Y10 ($Re_{M0} = 10$),
and Y14 ($Re_{M0} = 3$).
The plasma beta and the Hall parameter of these models are $\beta_0 =
100$ and $X_0 = 0.4$.
\label{fig:bp-t-y}}
\end{center}
\end{figure}
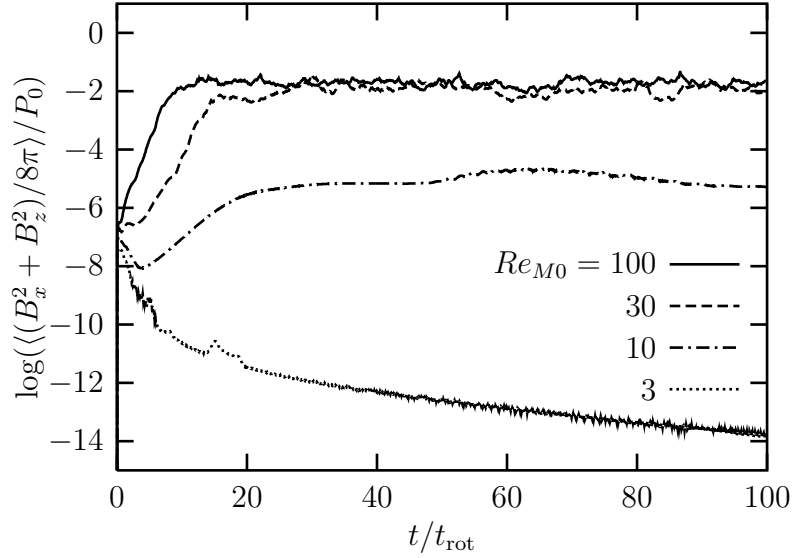

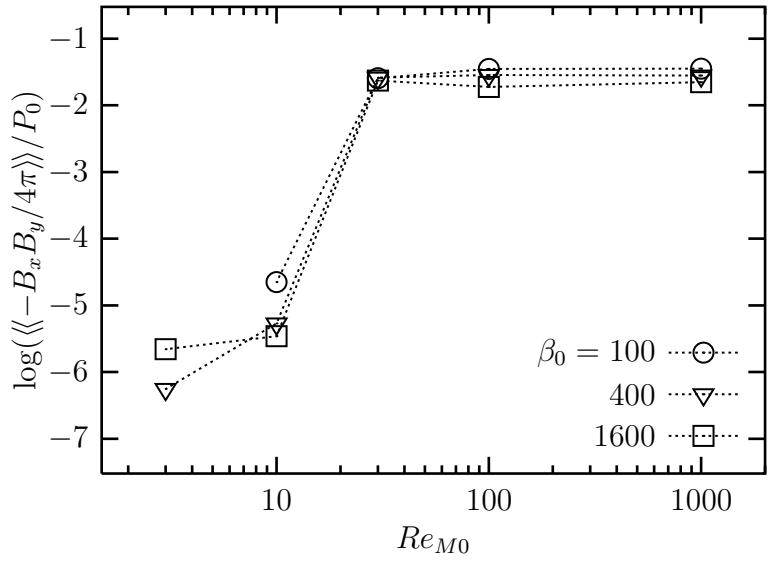
\begin{figure}
\begin{center}
\hspace*{-45pt}
\setlength{\unitlength}{0.1bp}%
{
}%
\begin{picture}(3600,2160)(0,0)%
{
}%
\put(3037,442){\makebox(0,0)[r]{\large{$1600$}}}%
\put(3037,599){\makebox(0,0)[r]{\large{$400$}}}%
\put(3037,756){\makebox(0,0)[r]{\large{$\beta_0=100$}}}%
\put(2200,50){\makebox(0,0){\large $Re_{M0}$}}%
\put(700,1180){%
\makebox(0,0)[b]{\shortstack{\large{$\log (\langle \negthinspace \langle - B_x B_y / 4 \pi \rangle \negthinspace \rangle / P_{0})$}}}%
}%
\put(3209,200){\makebox(0,0)[c]{\large $1000$}}%
\put(2409,200){\makebox(0,0)[c]{\large $100$}}%
\put(1609,200){\makebox(0,0)[c]{\large $10$}}%
\put(900,1940){\makebox(0,0)[r]{\large $-1$}}%
\put(900,1689){\makebox(0,0)[r]{\large $-2$}}%
\put(900,1437){\makebox(0,0)[r]{\large $-3$}}%
\put(900,1186){\makebox(0,0)[r]{\large $-4$}}%
\put(900,934){\makebox(0,0)[r]{\large $-5$}}%
\put(900,683){\makebox(0,0)[r]{\large $-6$}}%
\put(900,431){\makebox(0,0)[r]{\large $-7$}}%
\end{picture}%
\\ 
\caption
{Saturation level of the Maxwell stress as a function of the magnetic
Reynolds number $Re_{M0}$ for uniform $B_y$ models.
All the models are without the Hall effect ($X_0 = 0$), and the initial
field strength is $\beta_0 = 100$, 400, and 1600.
\label{fig:wm-rm-y}}
\end{center}
\end{figure}

\begin{figure}
\begin{center}
\hspace*{-45pt}
\setlength{\unitlength}{0.1bp}%
{
}%
\begin{picture}(3600,2376)(0,0)%
{
}%
\put(3037,442){\makebox(0,0)[r]{\large{$100$}}}%
\put(3037,599){\makebox(0,0)[r]{\large{$10$}}}%
\put(3037,756){\makebox(0,0)[r]{\large{$0.4$}}}%
\put(3037,913){\makebox(0,0)[r]{\large{$0.2$}}}%
\put(3037,1070){\makebox(0,0)[r]{\large{$X_{0} = 0$}}}%
\put(2225,2325){\makebox(0,0){\large $Re_{M0}' \equiv c_{s0}^2/\eta \Omega$}}%
\put(2225,50){\makebox(0,0){\large $Re_{M0} \equiv v_{A0}^2/\eta \Omega$}}%
\put(700,1188){%
\makebox(0,0)[b]{\shortstack{\large{$\log (\langle \negthinspace \langle (B_x^2 + B_z^2)/ 8 \pi \rangle \negthinspace \rangle / P_{0})$}}}%
}%
\put(3214,2176){\makebox(0,0)[c]{\large $100000$}}%
\put(2430,2176){\makebox(0,0)[c]{\large $10000$}}%
\put(1646,2176){\makebox(0,0)[c]{\large $1000$}}%
\put(3214,200){\makebox(0,0)[c]{\large $1000$}}%
\put(2430,200){\makebox(0,0)[c]{\large $100$}}%
\put(1646,200){\makebox(0,0)[c]{\large $10$}}%
\put(950,2076){\makebox(0,0)[r]{\large $0$}}%
\put(950,1854){\makebox(0,0)[r]{\large $-2$}}%
\put(950,1632){\makebox(0,0)[r]{\large $-4$}}%
\put(950,1410){\makebox(0,0)[r]{\large $-6$}}%
\put(950,1188){\makebox(0,0)[r]{\large $-8$}}%
\put(950,966){\makebox(0,0)[r]{\large $-10$}}%
\put(950,744){\makebox(0,0)[r]{\large $-12$}}%
\put(950,522){\makebox(0,0)[r]{\large $-14$}}%
\put(950,300){\makebox(0,0)[r]{\large $-16$}}%
\end{picture}%
\\ 
\caption
{Saturation level of the poloidal component of the magnetic energy as a
function of the magnetic Reynolds number $Re_{M0}$ for uniform $B_y$
models ($\beta_0 = 100$).
Open circles denote the models with only the ohmic dissipation ($X_0 =
0$), and the other marks are including also the Hall effect ($X_0 =
0.2$, 0.4, 10, and 100).
\label{fig:bp-rm-y}}
\end{center}
\end{figure}
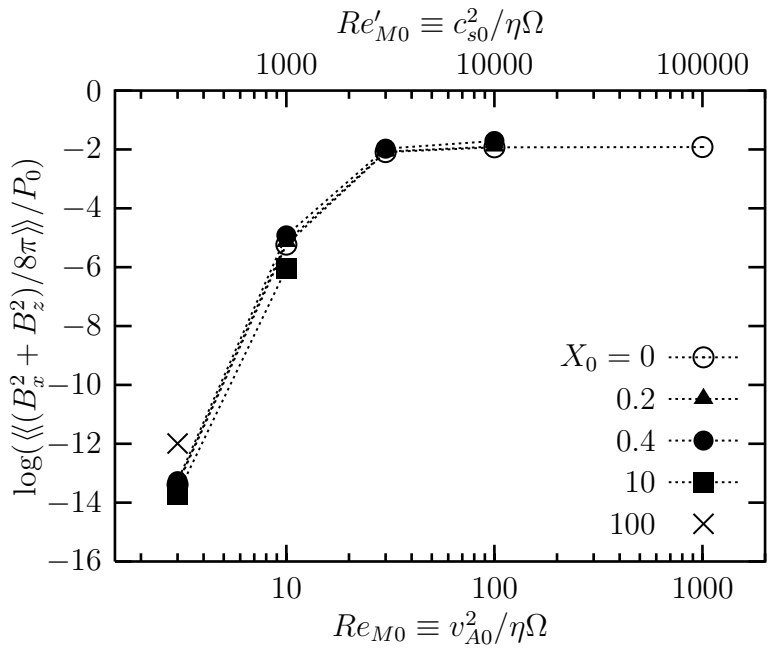

\begin{figure}
\begin{center}
\hspace*{-45pt}
\setlength{\unitlength}{0.1bp}%
{
}%
\begin{picture}(3600,2160)(0,0)%
{
}%
\put(2200,50){\makebox(0,0){\large $\log \langle \negthinspace \langle v_{Az}^2/\eta \Omega \rangle \negthinspace \rangle$}}%
\put(700,1180){%
\makebox(0,0)[b]{\shortstack{\large{$\log \alpha \equiv \log (\langle \negthinspace \langle w_{xy} \rangle \negthinspace \rangle / P_{0})$}}}%
}%
\put(3450,200){\makebox(0,0)[c]{\large $6$}}%
\put(3065,200){\makebox(0,0)[c]{\large $4$}}%
\put(2681,200){\makebox(0,0)[c]{\large $2$}}%
\put(2296,200){\makebox(0,0)[c]{\large $0$}}%
\put(1912,200){\makebox(0,0)[c]{\large $-2$}}%
\put(1527,200){\makebox(0,0)[c]{\large $-4$}}%
\put(1142,200){\makebox(0,0)[c]{\large $-6$}}%
\put(900,2060){\makebox(0,0)[r]{\large $1$}}%
\put(900,1864){\makebox(0,0)[r]{\large $0$}}%
\put(900,1669){\makebox(0,0)[r]{\large $-1$}}%
\put(900,1473){\makebox(0,0)[r]{\large $-2$}}%
\put(900,1278){\makebox(0,0)[r]{\large $-3$}}%
\put(900,1082){\makebox(0,0)[r]{\large $-4$}}%
\put(900,887){\makebox(0,0)[r]{\large $-5$}}%
\put(900,691){\makebox(0,0)[r]{\large $-6$}}%
\put(900,496){\makebox(0,0)[r]{\large $-7$}}%
\put(900,300){\makebox(0,0)[r]{\large $-8$}}%
\end{picture}%
\\
\caption
{Saturation level of the stress $\langle \negthinspace \langle w_{xy}
\rangle \negthinspace \rangle / P_0$ as a function of the magnetic
Reynolds number $\langle \negthinspace \langle v_{{\rm A}z}^2 / \eta
\Omega \rangle \negthinspace \rangle$ at the nonlinear stage.
Circles, triangles, and squares denote the models started with a uniform
vertical field, zero-net flux vertical field, and a uniform toroidal
field, respectively.
Filled marks means the results of the models including the Hall term,
and open marks are without the Hall effect.
Crosses are the results obtained by Fleming et al. (2000).
Upper and lower solid arrows indicate the averages of the ideal MHD runs
in Hawley et al. (1995) for initially uniform $B_z$ and initially
uniform $B_y$, respectively.
\label{fig:alpha-rmz}}
\end{center}
\end{figure}
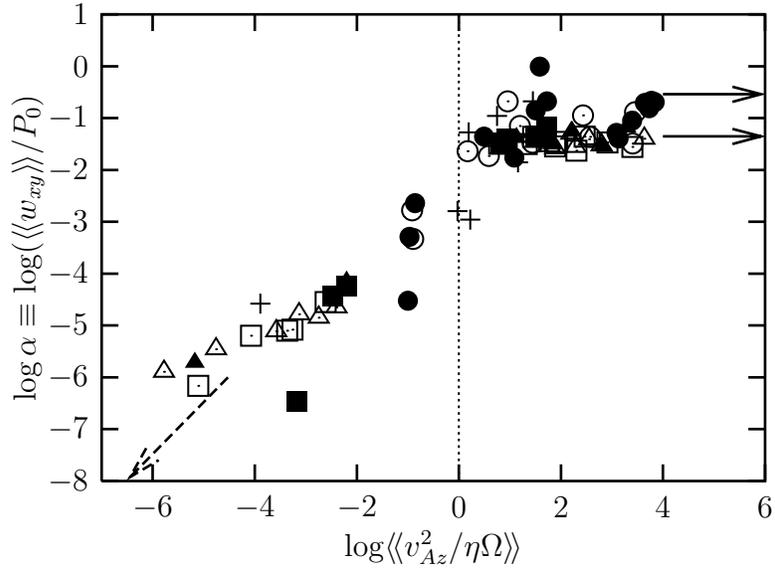

\end{document}